\documentclass[%
 reprint,
 amsmath,amssymb,
 aps,
prb,
floatfix,
]{revtex4-2}

\usepackage{graphicx}
\usepackage{dcolumn}
\usepackage{bm}
\usepackage{hyperref}
\usepackage{xcolor}
\usepackage[mathlines]{lineno}

\DeclareMathOperator{\tr}{tr}

\begin{document}

\preprint{APS/123-QED}

\title{Computation of High Frequency Magnetoelastic Waves in Layered
  Materials}

\author{Samuel J. Ryskamp}
\author{Mark A. Hoefer} \email{hoefer@colorado.edu}
\affiliation{%
Department of Applied Mathematics \\
University of Colorado Boulder
}%

\date{\today}

\begin{abstract}
  The direct calculation of magnetoelastic wave dispersion in layered   media is presented using an efficient, accurate computational   technique.  The governing, coupled equations for elasticity and   magnetism, the Navier and Landau-Lifshitz equations, respectively,   are linearized to form a quadratic eigenvalue problem that   determines a complex web of wavenumber-frequency dispersion branches   and their corresponding mode profiles.  Numerical discretization of   the eigenvalue problem via a spectral collocation method (SCM) is   employed to determine the complete dispersion maps for both a   single, finite-thickness magnetic layer and a finite   magnetic-nonmagnetic double-layer.  The SCM, previously used to   study elastic waves in non-magnetic media, is fast, accurate, and   adaptable to a variety of sample configurations and geometries.   Emphasis is placed on the extremely high frequency regimes being   accessed in ultrafast magnetism experiments.  The dispersion maps   and modes provide insight into how energy propagates through the   coupled system, including how energy can be transferred between   elastic- and magnetic-dominated waves as well as between different   layers.  The numerical computations for a single layer are further   understood by a simplified analytical calculation in the   high-frequency, exchange-dominated regime where the resonance   condition required for energy exchange (an anticrossing) between   quasi-elastic and quasi-magnetic dispersion branches is determined.   Nonresonant interactions are shown to be well approximated by the   dispersion of uncoupled elastic and magnetic waves.  The methods and   results provide fundamental theoretical tools to model and   understand current and future magnetic devices powering spintronic   innovation.
\end{abstract}

\maketitle

\author{Samuel J. Ryskamp\thanks{ Department of Applied Mathematics, University 
of Colorado, Boulder, CO (samuel.ryskamp@colorado.edu)} \and Mark A. Hoefer\thanks{ Department of Applied Mathematics, University 
of Colorado, Boulder, CO} }

\date{October 2021}

\maketitle

\section{Introduction}

The coupling between a material's vibrational and spin degrees of
freedom is a fundamental feature of magnetic materials.  While the
existence of this coupling has been known for over a century
\cite{richardson_mechanical_1908,barnett_magnetization_1909,einstein_1915,einstein_1915b,barnett_1915},
most theoretical research has focused on the low frequency regime
\cite{kamra_2015_elastic_excitation,streib_damping_2018,ruckriegel_long-range_2020,vanderveken_confined_2021}
while analytical studies emphasize simple geometries such as the bulk
\cite{kittel_1958,gurevich_1996} or thin film
\cite{vanderveken_confined_2021} limits.  Many of the magnetic devices
that enable current and future spintronic/magnonic applications are
multilayer stacks of magnetic and nonmagnetic materials
\cite{2020_mag_roadmap,barman_2021_2021,chumak_advances_2022}.
Furthermore, the ultrashort optical or acoustic pulses being used to
demagnetize samples in ultrafast magnetism
\cite{kirilyuk_ultrafast_2010,walowski_perspective_2016} or to
approach the plastic limit of a solid in ultrafast magnetoacoustics
\cite{vlasov_modern_2022}, respectively, are known to excite the
extremely high frequency (EHF) band ($30$--$300$ GHz)
\cite{zhang_high-frequency_2020,maznev_generation_2021,turenne_nonequilibrium_2022}.


Magnetoelastic coupling
\cite{kittel_1958,weiler_2011_resonance,kamra_2015_elastic_excitation,holanda_conversion_2018,an_momentum_2020,vanderveken_confined_2021}
and layering effects on the dispersion of purely elastic waves
\cite{Lowe_1995,quintanilla_2015_spectral,quintanilla_2015,ramasawmy_elastic_mat},
separately, are active areas of research, but few studies exist that
combine them. Although much of the original theoretical work on
magnetoelastic coupling is many decades old
\cite{becker_doring,kittel_1949,herring_kittel_1951,kittel_1958,schlomann_phonons_1960,soohoo_1960,callen_magnetoelastic_1963,camley_1979},
recent technological advances have revived and expanded interest in
magnetoelastic waves. At the heart of this growth is the drive to
enable energy transfer between phonons and magnons.  For example,
recent research has studied magnetic spin currents in thin layers,
particularly with applications to spintronics and magnonics
\cite{2020_mag_roadmap,barman_2021_2021,chumak_advances_2022}. This
interest has led to experimental advances in the study of
magnetoelastic interactions
\cite{weiler_2011_resonance,holanda_conversion_2018,spin_mech_quant,spin_phonon_2019,an_momentum_2020,godejohann_magnon_2020},
in part with the goal of generating spin waves by mechanical motion
and vice-versa \cite{matsuo_2015,kamra_2015_elastic_excitation}.
Other recent applications of magnetoelastic effects include
computation \cite{sadovnikov_computation_2019}, logic structures
\cite{verba_nonreciprocity_2019}, antennas \cite{yao_antennas_2020},
and smart materials \cite{fohtung_magnetostriction_2021}.

Similarly, there are many physical applications for calculating the
dispersion of elastic waves in a layered material, all of which are
made more general by the introduction of magnetic coupling. One
application is nondestructive evaluation, where elastic waves are
excited in a material in order to identify defects or other internal
properties \cite{cheeke_2012,kundu_2019,ramasawmy_elastic_mat}. If a
material is magnetic, one must take this property into account
\cite{fedders_acoustic_1974, wang_infrastructure_2014}. Elastic waves
in layered materials arise naturally in the geophysical sciences, such
as in the propagation of seismic waves
\cite{cheeke_2012,pak_2002,roganov_2014}. Some have found value by
incorporating magnetic effects there as well
\cite{chattopadhyay_selfreinforced_2010}.

Past analytical studies of magnetoelastic waves have primarily focused
on either the bulk $d \to \infty$ or thin film $kd \ll 1$ limits,
where $k$ is the angular wavenumber and $d$ is the material thickness
\cite{kittel_1958,gurevich_1996,vanderveken_confined_2021}. These
limits admit exact analytical solutions for the dispersion curves,
which provide important insight into the nature of phonon-magnon
energy conversion. Application of a phenomenological spin wave
decomposition has also recently been proposed
\cite{verba_nonreciprocal_2018}. The finite thickness regime, where
boundary effects must be considered, is more complicated and largely
unstudied analytically \cite{kamra_2015_elastic_excitation}.

A traditional method for determining the dispersion of elastic waves
in layered media is the general matrix method (GMM), which involves a
complex exponential profile ansatz in the direction of layering
\cite{knopoff_1964,nayfeh_1991,Lowe_1995,ramasawmy_elastic_mat}. A
previous numerical study of magnetoelastic waves utilized a GMM-type
approach to trace individual magnetoelastic dispersion curves for
surface waves in a magnetic film on a semi-infinite substrate
\cite{camley_1979}.  However, the complexity of the magnetoelastic
equations makes the GMM approach impractical for multiple, finite
thickness layers.  Another, very recent study, utilized
post-processing of coupled elastic-micromagnetic, time-dependent
simulations to determine the dispersion curves of a single layer
magnetoelastic wave guide by extracting space-time frequency curves
for certain configurations of initial, boundary data
\cite{vanderveken_confined_2021}. Both of these approaches are quite
computationally intensive. To our knowledge, the full dispersion map
of a magnetoelastic, layered material has never been determined
directly.

Despite the typical layered geometry of magnetic devices and the high
frequencies now being accessed, there are few developed theoretical
tools for investigating magnetoelastic wave propagation and little in
the way of comprehensive descriptions for these common configurations
and operating regimes.  In this paper, we directly compute
wavenumber-frequency dispersion maps and the corresponding vertical
mode profiles of magnetoelastic, layered materials using the spectral
collocation method (SCM), an accurate, fast numerical method to
compute dispersion maps and mode profiles for single and multiple,
finite thickness magnetic and elastic layers. A recent innovation for
calculating elastic dispersion in nonmagnetic materials, the SCM
utilizes a discretization at special points (zeros of Chebyshev
polynomials) and differentiation matrices for derivatives in the
direction perpendicular to the layering in order to achieve efficient,
accurate computation
\cite{adamou_2004,quintanilla_2015,quintanilla_2015_spectral}. The
linearized magnetoelastic equations result in a quadratic eigenvalue
problem. For a fixed wavenumber, each eigenvalue corresponds to a
frequency branch and the corresponding eigenvector is an interpolant
for the vertical mode profiles of magnetic and elastic components in
each layer. The mode profiles are used to classify dispersion branches
in terms of the energy residing in elastic or magnetic components and
in which layer the energy is concentrated. Using the SCM, we calculate
and analyze full magnetoelastic dispersion maps for a
yttrium-iron-garnet (YIG) single layer, YIG on a nonmagnetic
gadolinium gallium garnet (GGG) substrate, and nickel (Ni) on a
nonmagnetic silicon nitride (Si$_3$N$_4$) substrate.  This method is
shown to be easy to implement and is rapidly convergent, with
computational times measured in minutes on a conventional laptop
computer.

For magnetoelastic boundary value problems, other discretization
methods such as traditional finite differences applied to $N$
equispaced grid points, exhibit errors that decrease algebraically
$\mathcal{O}(N^{-n})$ for some $n$.  Usually, $n = 2$ is used in
practice.  The SCM method benefits from its superior accuracy because
interpolation at the nonuniform Chebyshev points is almost optimal in
a certain sense, achieving errors that decrease faster than any power
of $N$ for smooth functions
\cite{trefethen_matlab,boyd_chebyshev_2001}.  We observe almost
exponential convergence $\mathcal{O}(e^{-N})$ in our computations (see
the validation study in App.~\ref{sec:validation-scm}), hence many
fewer grid points are required to obtain accurate results with the SCM
method than for traditional finite differences, providing significant
computational speedup.  About $N = 12$ grid points per layer is
sufficient for the regimes studied here to achieve high resolution.
Furthermore, the clustering of Chebyshev grid points near interfaces
ensures high resolution of surface modes and other interfacial mode
features.

Our numerical computations are complemented by an analytical
calculation in the EHF, exchange-dominated regime that is used to
explicitly determine magnetoelastic dispersion curves in a single,
finite thickness ferromagnetic layer. We find that crossings between
phonon and magnon dispersion branches can be resonant or nonresonant
in the presence of magnetoelastic coupling. If the vertical mode
profiles are of the same order, they are resonant and an anticrossing
appears.  Anticrossings result in the transfer of energy between
quasi-elastic and quasi-magnetic dispersion branches with a
well-defined gap width linearly proportional to the magnetoelastic
coupling.  If the waves are nonresonant, no anticrossing occurs, and
the dispersion branches intersect with very weak modification that is
proportional to the square of the magnetoelastic coupling.  Our
calculations also show that the magnetoelastic gap width decreases
with increasing frequency and decreasing wave number.


Our work is also motivated by recent experiments where an ultrafast
X-ray free electron laser (XFEL) was used to excite and measure
far-from-equilibrium conditions in layered magnetic samples
\cite{iacocca_2019_ultrafast,hagstrom_2021_ultrafast,turenne_nonequilibrium_2022}.
Because of this, we focus primarily on EHF, exchange-dominated
interactions with a perpendicular applied field.  However, these
choices are incidental to the applicability of the SCM approach. The
SCM is quite general and can be readily applied to similar problems
involving in-plane fields, lower frequencies, more complex
anisotropies, and even non-planar geometries. We will demonstrate some
of this versatility by presenting dispersion curves for multiple
materials and sample sizes in the EHF band as well as in the more
traditional single GHz regime.

The outline of this work is as follows.  After introducing the Navier
and Landau-Lifshitz equations for elasticity and magnetism in
Sec.~\ref{sec:mag_eq}, their linearization about a perpendicularly
magnetized, static configuration in Sec.~\ref{sec:line-magn-equat}
results in a quadratic eigenvalue problem in differential form.  An
asymptotic analysis of this eigenvalue problem, neglecting dipole
effects, for weak magnetoelastic coupling in Sec.~\ref{sec:analysis}
determines the resonance condition for anticrossings and phonon-magnon
energy transfer.  In Sec.~\ref{sec:scm}, we introduce the spectral
collocation method (SCM) and use it in Sec.~\ref{sec:results} to
compute dispersion maps for single and double layer films of various
thicknesses.  Finally, we conclude with a discussion in
Sec.~\ref{sec:disc-concl}.

\section{Governing equations}
\label{sec:mag_eq}

This section summarizes the equations for magnetoelastic waves mostly
following
\cite{vanderveken_confined_2021,kamenetskii_magnetoelastic_2021},
with additional material synthesized from other references including
\cite{kittel_1958,gurevich_1996}. Additional helpful sources regarding
magnetic and elastic waves include
\cite{becker_doring,kronmuller_2019,camley_1979,herring_kittel_1951,landau_lifshitz,harris_elastic,Liang_2014}.

\subsection{Energy density}
The energy density of a magnetoelastic material consists of the following components \cite{vanderveken_confined_2021}
\begin{equation}
    \label{eq:tot_en_density}
    \mathcal{E}_{\rm tot} =  \mathcal{E}_{\rm dip}+\mathcal{E}_{\rm ex} +\mathcal{E}_{\rm Z}  + \mathcal{E}_{\rm an}   + \mathcal{E}_{\rm el} + \mathcal{E}_{\rm kin}+ \mathcal{E}_{\rm c}.
\end{equation}
These terms represent, from left to right, the energy density due to
dipole (demagnetization) effects, exchange interactions, the Zeeman
effect (applied field), anisotropy energy, elastic potential, kinetic
energy, and magnetoelastic coupling. For a material with magnetization
vector $\bm{M}$, the first three energy density quantities are
calculated as
\begin{align}
    \mathcal{E}_{\rm dip} &= -\frac{\mu_0}{2}(\bm{M} \cdot \bm{H}_{\rm dip}), \\   \mathcal{E}_{\rm ex} &= \frac{A_{\rm ex}}{M_{\rm s}}|\nabla \bm{M}|^2, \\
    \mathcal{E}_Z &= -\mu_0(\bm{M} \cdot \bm{H}_0).
\end{align}
where $A_{\rm ex}$ is the exchange stiffness coefficient, $M_{\rm s}$
is the saturation magnetization of the material,
$\mu_0=4 \pi \times 10^{-7}$ H/m is the magnetic permeability of free
space, $\bm{H}_{\rm dip}$ is the induced field by the magnet, which is
computed here under the
magnetostatic approximation of Maxwell's equations by assuming that wave speeds are much smaller than the speed of light
(see eq.~\eqref{eq:poisson}),
and $\bm{H}_0$ is an applied external field. The intrinsic crystalline
anisotropy energy density $\mathcal{E}_{\rm an}$ depends on the
material's favored direction of magnetization.  For a magnetically
cubic material, this is calculated as \cite{kronmuller_2019}
\begin{equation*}
  \mathcal{E}_{an} = \frac{K_1}{M_s^2} \sum_{i\ne j} M_i^2 M_j^2 +
  \frac{K_2}{M_s} M_1^2 M_2^2 M_3^2 .
\end{equation*}
In the materials of interest here, the intrinsic crystalline
anisotropy is weaker than the other energy terms
\cite{weiler_2011_resonance,strauss_properties_1968}.  In the interest
of simplicity, we will ignore it but its incorporation into the
computation of magnetoelastic dispersion is straightforward because it
is a local, undifferentiated effective field term.

Under Hooke's Law, which assumes a linear relationship between stress and strain, the elastic energy density is calculated as \cite{love_elasticity_1906,kittel_1958}
\begin{equation}
    \label{eq:elastic_en}
    \mathcal{E}_{\rm el} =\frac{1}{2} \sum_{i,j} \sigma_{ij} \varepsilon_{ij} ,
\end{equation}
where $\sigma$ is the stress tensor and $\varepsilon$ is the strain
tensor. Kinetic energy for the displacement $\bm{u}$ is calculated as
\begin{equation}
    \mathcal{E}_{\rm kin} = \frac{\rho}{2} \left|\bm{u}_t \right|^2,
\end{equation}
where $\rho$ is the material density. 

The coupling energy for a cubic material is calculated as an expansion of anisotropy and elastic energy around the strain $\varepsilon$ and the magnetization $\bm{M}$ as \cite{kittel_1958} 
\begin{equation}
    \label{eq:coupling_energy}
    \begin{split}
      \mathcal{E}_{\rm c} = \frac{B_1}{M_{\rm
      s}^2}\Big(\varepsilon_{11}M_{1}^2 + \varepsilon_{22}M_2^2 +
      \varepsilon_{33} M_3^2 \Big)+\\+\frac{2B_2}{M_{\rm
      s}^2}\Big(\varepsilon_{12}M_1 M_2 + \varepsilon_{23}M_2
      M_3+\varepsilon_{31}M_1  M_3 \Big).\end{split}
\end{equation}
The coupling coefficients in an isotropic medium are defined as
$B_1 =-3\mu\lambda_{100}$ and $B_2 =-3\mu\lambda_{111}$, where $\mu$
is the Lam\'{e} coefficient and $\lambda_{100}$ and $\lambda_{111}$
are dimensionless coupling coefficients for a particular material.

The total energy in the magnetic material occupying the region
$U$ is calculated by a volume integral of the energy density,
\begin{equation}
\label{eq:en_tot}
    E = \int_U \mathcal{E}_{\rm tot} d\bm{x}.
\end{equation}
Equilibrium states of a magnetoelastic material are local minima of the total energy $E$ \eqref{eq:en_tot}.

\subsection{Magnetism}
The relaxation from a non-equilibrium state of the magnetization
vector $\bm{M}(\bm{x},t)$, defined within the region
$U \subset \mathbb{R}^3$ is governed by the Landau-Lifshitz-Gilbert
equation \cite{landau_lifshitz,gilbert_phenomenological_2004,bertotti_2009},
\begin{equation}
  \begin{split}
    \label{eq:full_landau_lifshitz}
    \frac{\partial \bm{M}}{\partial t} &= - |\gamma| \mu_0 \bm{M}
                                         \times \bm{H}_{\rm eff} +
                                         \frac{\alpha }{M_{\rm s}}
                                         \bm{M} \times \frac{\partial
                                         \bm{M}}{\partial t},
    \\
    \bm{H}_{\rm eff} &= -\frac{1}{\mu_0}\frac{\delta E}{\delta \bm{M}}.
  \end{split}
\end{equation}
where $\alpha$ is the nondimensional Gilbert damping parameter and
$|\gamma|=1.76 \times 10^{-11}$ rad/Ts is the gyromagnetic ratio for
an electron. One important property of \eqref{eq:full_landau_lifshitz}
is that the magnitude of $\bm{M}$ is conserved, and this magnitude is
called the saturation magnetization $M_{\rm s}$ of the material,
i.e. $|\bm{M}|=M_{\rm s}$. Consequently, it is often convenient to
refer to the normalized vector field $\bm{\zeta} = \bm{M}/M_{\rm s}$
so that $|\bm{\zeta}|=1$.

 The effective magnetic field $\bm{H}_{\rm eff}$ is calculated as the variational derivative of the total energy \eqref{eq:en_tot} with respect to magnetization \cite{garcia_2007}. Since there are five components of the energy \eqref{eq:tot_en_density} that depend on $\bm{M}$, the effective magnetic field can be decomposed into five corresponding components,
\begin{equation}
    \label{eq:defn_H}
    \bm{H}_{\rm eff} = \bm{H}_{\rm dip} + \bm{H}_{\rm ex}+ \bm{H}_0+\bm{H}_{\rm an}+\bm{H}_{\rm c}.
\end{equation}
The dipole field $\bm{H}_{\rm dip}$, also known as the stray field or the demagnetizing field, is obtained by solving Maxwell's equations \cite{garcia_2007,stancil_spin_waves_2009,harte_ripple_1968}, in which the magnetization induces a magnetic field. We will assume the magnetostatic approximation, which ignores the time variation of electric fields, reducing Maxwell's equations to
\begin{subequations}
\label{eq:poisson}
\begin{align}
    \label{eq:nabla_b}
    \nabla \cdot \bm{B} &= 0, \\
    \label{eq:nabla_h}
    \nabla \times \bm{H}_{\rm dip} &= 0,
\end{align}
where $\bm{B}$ is the total magnetic induction field
\begin{equation}
    \label{eq:def_b}
    \bm{B} = \mu_0(\bm{H}_{\rm dip} + \bm{M}).
\end{equation}
The appropriate boundary conditions are \cite{harte_ripple_1968}
\begin{align}
    \label{eq:full_bc_1}
  \bm{B}\cdot \bm{n} &= \mathrm{continuous}, \qquad \bm{x} \in
                       \partial U, \\ 
    \label{eq:full_bc_2}
    \bm{H}_{\rm dip} \times \bm{n} &= \mathrm{continuous}, \qquad
                                     \bm{x} \in \partial U,\\ 
    \label{eq:full_bc_3}
    \lim_{|\bm{x}|\to\infty} \bm{H}_{\rm dip} &= 0, \qquad \qquad
                                                \qquad \, \bm{x} \in
                                                U^C \backslash
                                                \partial U,
\end{align}\end{subequations}
where $\partial U$ is the boundary of $U$.  The exchange
field is due to the property that ferromagnetic materials tend to
align spins along a common direction \cite{garcia_2007} and is given
by
\begin{equation}
\label{eq:exchange_full}
    \bm{H}_{\rm ex} = \ell^2 \Delta \bm{M},
\end{equation}
where $\ell$ is the exchange length of the material defined as $\ell^2 = 2 A_{\rm ex}/(\mu_0 M_{\rm s}^2)$. Physically, $\ell$ represents the length scale where exchange effects are dominant. For lengths much larger than $\ell$, dipole effects dominate. 

Although the materials considered here are magnetically cubic, we will assume that the effect of this anisotropy on the total effective field is small enough so that it can be ignored. Finally, the effective field due to magnetoelastic coupling can be calculated from the coupling energy density \eqref{eq:coupling_energy} as 
\begin{equation}
    \label{eq:couple_eff_field}
    \bm{H}_{\rm c} = -\frac{2 }{\mu_0 M_{\rm s}} \begin{bmatrix}B_1\varepsilon_{11} \zeta_1+B_2(\varepsilon_{12}\zeta_2 + \varepsilon_{13}\zeta_3) \\ B_1 \varepsilon_{22}\zeta_2 +B_2(\varepsilon_{12}\zeta_1 + \varepsilon_{23}\zeta_3) \\ B_1\varepsilon_{33}\zeta_3 + B_2 (\varepsilon_{13}\zeta_1 + \varepsilon_{23}\zeta_2) \end{bmatrix}.
\end{equation} 
The boundary conditions for magnetism are free spin, i.e.
\begin{equation}
\label{eq:mag_gen_bc}
    \nabla_{\bm{n}} \cdot \bm{M} = 0, \qquad \bm{x} \in \partial U,
\end{equation}
where $\bm{n}$ is the unit vector normal to the surface, and
$\nabla_{\bm{n}} \cdot \bm{M}$ represents the divergence in the
direction of $\bm{n}$.

Recently, inertial effects have been observed in magnetic samples
driven at 0.4--1.0 THz frequencies \cite{neeraj_inertial_2021}.
Inertia drives spin nutation corresponding to the relaxation of
angular momentum.  Inertia can be modeled by the addition of a second
derivative in time term proportional to $\alpha \tau/T$ to the
Landau-Lifshitz-Gilbert equation \eqref{eq:full_landau_lifshitz}
\cite{olive_beyond_2012,giordano_derivation_2020}.  Here, $\tau$ is
the material-dependent angular momentum relaxation time---estimated
and recently measured for NiFe and CoFeB \cite{neeraj_inertial_2021},
to be in the 1--100 ps range---and $T$ is the precessional
timescale---about 10 ps--1 ns here. Inertial effects may be relevant
when the effective field contribution, scaling with the nondimensional
parameter $\alpha \tau/T$, is comparable to other effective field
terms of interest.  For the materials in the present study, we
estimate $\alpha \tau/T$ to be no larger than $10^{-3}$ (and much
smaller in many cases) whereas the nondimensional magnetoelastic
coupling strength is $\epsilon \approx 0.02$ (see
eq.~\eqref{eq:epsilon}).  Since the EHF regime studied here is below
the THz frequency range and magnetoelastic coupling is sufficiently
strong in the materials studied, we neglect inertia
\cite{giordano_derivation_2020}.

\subsection{Elasticity}
The elastic displacement $\bm{u}$ in the linear regime is governed by the equation 
\begin{equation}
    \label{eq:full_elastic}
\rho \bm{u}_{tt} = \bm{f}, \quad \bm{f} = \nabla \cdot \sigma=\nabla \cdot  \frac{\delta E}{\delta \varepsilon}.
\end{equation}
With constant temperature, the variational derivative of energy is the
stress tensor $\frac{\delta E}{\delta \varepsilon} = \sigma$. We
assume an isotropic elastic material and neglect acoustic damping.
Then, Hooke's law for the total stress $\sigma$ in terms of the strain
$\varepsilon$ depends on two parameters,
\[ \sigma = \lambda \tr{(\varepsilon)}I + 2 \mu \varepsilon.\] The
constants $\lambda$ and $\mu$ are the Lam\'{e} moduli with units of
stress. For small displacements, we can in general approximate the
total strain tensor $\varepsilon$ as
\[
\varepsilon_{ij} = (\partial_j u_i + \partial_i u_j)/2.
\]

 Combining the above equations, we obtain a forced Navier equation for homogeneous, linear, isotropic magnetoelastic deformations
\begin{equation}
    \label{eq:navier}
       (\lambda + 2 \mu) \nabla (\nabla \cdot \bm{u}) - \mu \nabla \times \nabla \times \bm{u} + \bm{f}(\bm{m}) = \rho \ddot{\bm{u}},
\end{equation}
where $\bm{f}$ arises from the magnetoelastic coupling energy
$E_{\rm c}$.  When $\bm{f}\equiv \bm{0}$ (i.e. with no magnetic
effects present), by a Helmholtz decomposition we can recover the two
bulk speeds of sound for an isotropic material \cite{harris_elastic}, the
shear speed $c_S$ and the longitudinal speed $c_L$
\begin{equation}
  \label{eq:speeds of sound}
  c_S = \sqrt{\mu/\rho}, \quad c_L = \sqrt{(\lambda +  2\mu)/\rho} ,
\end{equation}
subject to the ordering $c_S < c_L$.  The variational derivative of
the coupling energy $E_{\rm c}$ yields the magnetic component of the
stress $\frac{\delta E_{\rm c}}{\delta \varepsilon} = \sigma^{\rm
  c}$. The divergence of the magnetic stress
$\bm{f}=\nabla \cdot \sigma^{\rm c}$ is the forcing term in the Navier
equation,
\begin{equation}
    \label{eq:mag_force}
{\small \bm{f} =  2B_1 \begin{bmatrix} \zeta_1 \frac{\partial \zeta_1}{\partial x_1} \\ \zeta_2 \frac{\partial \zeta_2}{\partial x_2} \\ 
    \zeta_3 \frac{\partial \zeta_3}{\partial x_3}\end{bmatrix} + B_2 \begin{bmatrix} \zeta_1 \left(\frac{\partial \zeta_2}{\partial x_2} + \frac{\partial \zeta_3}{\partial x_3} \right)+ \zeta_2 \frac{\partial \zeta_1}{\partial x_2}+ \zeta_3 \frac{\partial \zeta_1}{\partial x_3} \\
    \zeta_2 \left(\frac{\partial \zeta_1}{\partial x} + \frac{\partial \zeta_3}{\partial x_3} \right)+ \zeta_1 \frac{\partial \zeta_2}{\partial x_1}+ \zeta_3 \frac{\partial \zeta_2}{\partial x_3} \\
    \zeta_3 \left(\frac{\partial \zeta_1}{\partial x_1} + \frac{\partial \zeta_2}{\partial x_2} \right)+ \zeta_1 \frac{\partial \zeta_3}{\partial x_1}+ \zeta_2 \frac{\partial \zeta_3}{\partial x_2}
    \end{bmatrix}}.
\end{equation}
For an elastic material, the boundary conditions at the surface are
\begin{equation}
\label{eq:el_gen_bc}
    \bm{f}_{\rm surf} = \sigma \bm{n}, \qquad \bm{x} \in \partial U.
\end{equation}
where $\bm{f}_{\rm surf}$ is the traction force per unit surface (or
stress) at the boundary.

\section{Linearized magnetoelastic equations}
\label{sec:line-magn-equat}


\subsection{Physical assumptions}
 
\begin{figure}
    \centering
    \includegraphics[scale=.34]{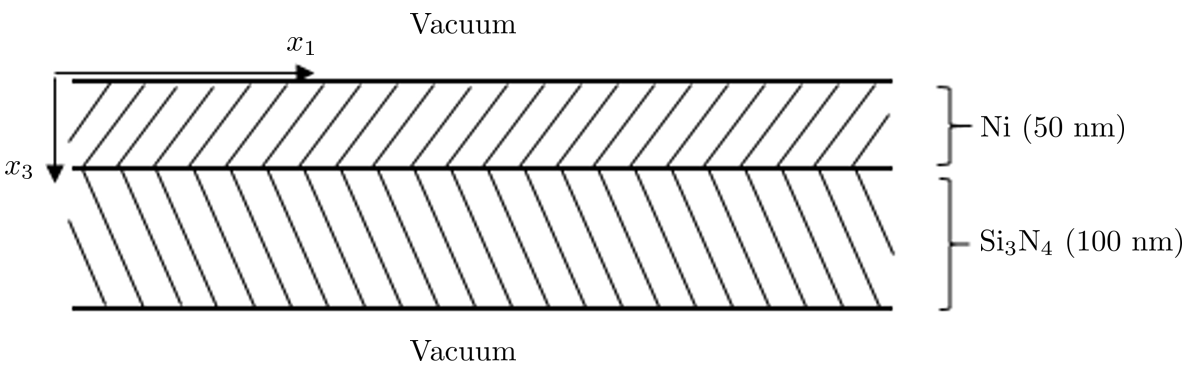}
    \caption{The experimental medium for which we are calculating the magnetoelastic dispersion relation: 100 nm of silicon nitride (Si$_3$N$_4$), a nonmagnetic substrate, supporting 50 nm of nickel, surrounded by a vacuum. For modeling purposes, the material is considered to be infinite in the $x_1$ and $x_2$ directions.}
    \label{fig:layered_material}
\end{figure}
We consider two magnetic and nonmagnetic materials.  They are chosen
for their use in recent experiments, their low magnetic damping in the
case of YIG, and their relatively large magnetoelastic coupling.  One
of the double layers we study consists of 100 nm of Si$_3$N$_4$
supporting a 50 nm
layer of nickel. See Fig.~\ref{fig:layered_material} for a
schematic. Besides this material sample, we will also study a 30 nm
single layer of YIG and both 30 nm/50 nm, 200 nm/300 nm YIG/GGG double
layers. YIG has been classically studied in the context of
magnetoelastic interactions due to its very low magnetic damping
(cf.~\cite{camley_1979,gurevich_1996}), and GGG is a natural substrate
for YIG \cite{camley_1979}. YIG and GGG have the added advantage of
comparable shear speeds of sound, which will simplify the
visualization of their dispersion relations. The material parameters
are shown in Table~\ref{table:speeds}.
\begin{table}[ht]
\centering
\begin{tabular}{|c || c | c| c | c |} 
 \hline
 & Ni \cite{crc_handbook} & Si$_2$N$_3$ \cite{azom_sin} & YIG \cite{gurevich_1996,strauss_properties_1968} & GGG \cite{graham_ggg_elastic_1970} \\
 \hline\hline
$c_L$ (m/s) & 6040  & 8000 & 7200 & 6400 \\ \hline
$c_S$ (m/s) & 3000 & 4500 & 3800 & 3500 \\ \hline
$\rho$ (kg/m$^3$) & 8900 & 3250 & 5170 & 7080 \\ \hline
$B_1$ (MJ/m$^3$) & 9.38 \cite{tian_2009} & X & 3.5 & X \\ \hline
$B_2$ (MJ/m$^3$) & 10 \cite{tian_2009} & X & 7 & X \\ \hline
$\ell$ (nm)  & 7.72 \cite{vanderveken2020} & X & 17.3 & X \\ \hline
$M_{\rm s}$ (kA/m)  & 480 \cite{vanderveken2020}& X & 140 & X \\
\hline
\end{tabular}
\caption{Parameter values for the four materials studied in this
  paper. The parameter values for Si$_2$N$_3$ were chosen from the
  ranges given by \cite{azom_sin} in order to minimize the shear
  speed, allowing for easier visualization.}
\label{table:speeds}
\end{table}

In this paper, we assume that all materials are elastically isotropic,
and that the magnetic lattice is cubic.  We also assume that the
material is infinitely wide in the transverse $x_1$ and $x_2$
directions, the displacements are small and therefore linear, and the
waves of interest propagate in the $x_1$-direction. We will limit
ourselves to materials with either a single magnetic layer, or a
single magnetic layer on a nonmagnetic substrate, with the magnetic
layer in both cases vertically localized to $0<x_3<d$.

The nondimensional Gilbert damping parameter $\alpha$ is as low as
$3 \times 10^{-5}$ for YIG \cite{camley_nonlinear_2011}.  Nickel is a
metallic ferromagnet where $\alpha$ was measured to be $0.024$
\cite{schoen_magnetic_2017}.  The nondimensional magnetoelastic
coupling strength is estimated to be $\epsilon \approx 0.02$ for YIG
and Ni (see eq.~\eqref{eq:epsilon}) so is much stronger than damping
in YIG.  Our focus here is on the real frequency response due to
magnetoelastic coupling of layered devices in the EHF regime, hence we
will neglect magnetic damping for our analysis, which is a reasonable
physical assumption for YIG and transition metal alloys.  Therefore,
all but one presented computational example is for YIG.  We compute
the magnetoelastic dispersion of Ni/SiN while neglecting damping, even
though its strength is comparable to magnetoelastic coupling, in order
to convey the robustness of the SCM method to material properties.
Damping introduces a nonzero imaginary part to the frequency
corresponding to wave attenuation and a nonzero frequency linewidth
\cite{camley_1979}.  Large enough damping could smear out the resonant
frequency anticrossings predicted here.

We assume that a magnetic field $\bm{H}_0 $ is applied in the
$x_3$-direction, perpendicular to the plane of the material, so that
$\bm{H}_0= \begin{bmatrix} 0 & 0 & H_0\end{bmatrix}^T$. The magnetic
field is sufficiently strong that, in the absence of waves, the
magnetization is saturated in the $x_3$-direction. This occurs when
$H_0 \geq M_{\rm s}$. In order to ensure saturation magnetization in
the vertical direction, we choose $\mu_0 H_0 = 0.25$ T for the
calculations involving YIG, and $\mu_0 H_0 = 0.65$ T for the
calcluations involving nickel.

It is important to stress that nearly all of these assumptions can be
relaxed without reducing the effectiveness of the SCM approach, which
is very versatile. SCM can be readily generalized to situations with
other geometries, more complex elastic or magnetic anisotropies
\cite{quintanilla_2015_spectral}, and to other physical effects such
as magnetic/acoustic damping and spin inertia.  What we focus on here
is a minimal but realistic model of the physics of layering and
magnetoelastic coupling in the ultrafast regime.

\subsection{Elastic equations}
We look for linear wave solutions of the form
\begin{equation}
    \label{eq:spec_ansatz}
    \bm{u}(\bm{x},t) = \begin{bmatrix} A_1(x_3) \\ A_2(x_3) \\ A_3(x_3) \end{bmatrix}
    e^{i(k x_1 - \omega t)}.
\end{equation}
Inserting \eqref{eq:spec_ansatz} into \eqref{eq:navier} and dividing out the common exponential factor yields, after substituting in the speeds \eqref{eq:speeds of sound},
\begin{subequations}
\label{eq:24}
\begin{align}
  -\omega^2 A_1 &= -c_L^2 k^2 A_1 + c_S^2 A_1'' + (c_L^2-c_S^2) i k
                  A_3' + \frac{f_1}{\rho}, \\
  -\omega^2 A_2 &= c_S^2 (A_2''-k^2 A_2) + \frac{f_2}{\rho}, \\
  -\omega^2 A_3 &= c_L^2  A_3'' -  c_S^2 k^2 A_3'' + (c_L^2-c_S^2) i k
                  A_1' + \frac{f_3}{\rho},
\end{align}\end{subequations}
where a prime denotes differentiation with respect to $x_3$.  Here,
$f_i/\rho$ should also be understood to be divided by the common
exponential factor.

\subsection{Magnetic equations}
For small oscillations, we linearize the magnetic waves around the
vertical saturation magnetization as \begin{equation}
\begin{split}
    \label{eq:cam_mag_ansatz}
    \bm{M}(\bm{x},t)&=\bm{M}_0+\bm{m}(\bm{x},t), \\ &= 
    \begin{bmatrix}0 \\ 0 \\ M_{\rm s}  \end{bmatrix} + 
    \begin{bmatrix}m_1(x_3) \\ m_2(x_3) \\ 0 \end{bmatrix} e^{i(k x_1 -\omega t )},
    \end{split}
\end{equation}
where $m_1$, $m_2\ll M_{\rm s}$. Since we are only interested in real dispersion relations for $\omega$ and $k$, we will neglect the damping in \eqref{eq:full_landau_lifshitz} to obtain an undamped Landau-Lifshitz equation, also known as the Larmor torque equation,
\begin{equation}
    \label{eq:landau_lifshitz}
    \frac{\partial \bm{M}}{\partial t} = - \gamma \mu_0 \bm{M} \times \bm{H}_{\rm eff}.
\end{equation} 
Given the linearization \eqref{eq:cam_mag_ansatz}, the exchange field \eqref{eq:exchange_full} has the form
$ \bm{H}_{\rm ex} = \ell^2 \left(\bm{m}''-k^2 \bm{m} \right)$.

In order to determine the field due to dipole effects
$\bm{H}_{\rm dip}$ we must first solve the magnetostatic Maxwell
equations \eqref{eq:poisson}. Since the dipole field is curl free, it
is the gradient of a scalar
\begin{equation}
  \label{eq:10}
  \bm{H}_{\rm dip} =
  \begin{cases}
    -\nabla \phi^+, & d < x_3, \\
    -\nabla \phi, & 0 < x_3 < d, \\
    -\nabla \phi^-, & x_3 < 0 .
  \end{cases}
\end{equation}
We follow \cite{harte_ripple_1968} to obtain a simplified 
equation for the dipole field.  In the absence of waves,
$\bm{M} = \bm{M}_0$ in \eqref{eq:cam_mag_ansatz} and the dipole field
\eqref{eq:10} with $\phi = \phi_0$ satisfying \eqref{eq:poisson} is
\begin{equation}
  \label{eq:11}
  \phi_0 = M_{\rm s} x_3, \quad \phi_0^+ = M_{\rm s} d, \quad \phi_0^-
  = 0 .
\end{equation}
For the wave-only term $\bm{M} = \bm{m}$ in
\eqref{eq:cam_mag_ansatz}, we seek a solution of \eqref{eq:poisson} in
the form
\begin{equation}
  \label{eq:12}
  \varphi = g(x_3) e^{i(kx_1 - \omega t)}, \quad \varphi^\pm =
  g^{\pm}(x_3) e^{i(kx_1 - \omega t)}
\end{equation}
for the dipole field \eqref{eq:10} with $\phi = \varphi$.  Then,
eq.~\eqref{eq:nabla_b} with \eqref{eq:def_b} imply
\begin{subequations}
  \label{eq:14}
\begin{align}
  \label{eq:13}
  -k^2 g(x_3) + g''(x_3) &= ik m_1(x_3), \\
  \label{eq:15}
  -k^2 g^{\pm}(x_3) + (g^\pm)''(x_3) &= 0 .
\end{align}
\end{subequations}
Equation \eqref{eq:15} with the boundary condition
\eqref{eq:full_bc_3} is solved with $g^\pm(x_3) = C_\pm e^{\mp k
  x_3}$. Since the second component of $\bm{H}_{\rm dip}$ is zero (no
$x_2$ dependence in \eqref{eq:10}), we can impose \eqref{eq:full_bc_2} by
taking $g(d) = C_+ e^{-kd}$, $g(0) = C_-$.  The boundary condition
\eqref{eq:full_bc_1} implies $g'(d) = -k C_+ e^{-kd}$, $g'(0) = k
C_-$.  Eliminating $C_\pm$ by combining these two conditions we
obtain
\begin{equation}
    \label{eq:g_equation}
    \begin{split}
      -k^2g(x_3)+g''(x_3) =  i k m_1(x_3),\\
      g'(d) = -k g(d), \qquad g'(0) = k g(0),
    \end{split}
\end{equation}
so that the complete dipole potential is $\phi = \phi_0 + \varphi$ and
the corresponding dipole field inside the magnet is
\begin{equation}
\label{eq:dip_eqn}
\bm{H}_{\rm dip} = -\nabla \phi =  \begin{bmatrix}- i k g(x_3) e^{i(kx_1-\omega t)} \\ 0 \\ -g'(x_3)e^{i(kx_1-\omega t)} -M_{\rm s} \end{bmatrix},
\end{equation}
Note that no averaging assumptions were made for the dipole field
\eqref{eq:dip_eqn}, which is an exact solution to the magnetostatic
Maxwell equations \eqref{eq:poisson}.  Although
eq.~\eqref{eq:g_equation} can be solved using Green's function
techniques \cite{kalinikos_theory_1986}, we will obtain it numerically
as part of the SCM.

\subsection{Coupling}
From the magnetic wave ansatz \eqref{eq:cam_mag_ansatz} we can approximate $\zeta_3 \approx 1$, $\zeta_2 \approx \zeta_1 \approx 0$, $\partial \zeta_3 / \partial x_i \approx 0$. The magnetoelastic forcing terms from \eqref{eq:mag_force} simplify to 
\begin{equation}
    \label{eq:magnetostrict_force}
    \bm{f}=  \frac{ B_2}{M_{\rm s}}\begin{bmatrix}  m_1' \\ m_2' \\i k m_1     \end{bmatrix}e^{i(kx_1-\omega t)}. 
\end{equation}
  The effective field due to elastic coupling from \eqref{eq:landau_lifshitz} and \eqref{eq:defn_H} also simplifies to become
 \begin{equation}
    \bm{H}_{\rm c}  = - \frac{1}{\mu_0 M_{\rm s}}\begin{bmatrix}
     B_2 ( A_1' + i k A_3) \\
     B_2 A_2' \\ 2B_1 A_3'
    \end{bmatrix}e^{i(kx_1-\omega t)}.
    \end{equation}
    Inserting \eqref{eq:magnetostrict_force} into \eqref{eq:24} and \eqref{eq:defn_H} into \eqref{eq:landau_lifshitz} and neglecting quadratic terms yields
    \begin{widetext}

 \begin{subequations}
\label{eq:full_coupled_system} 
\begin{align}
    -\omega^2 A_1 &= -c_L^2 k^2 A_1 + c_S^2 A_1'' + (c_L^2-c_S^2) i k A_3' + \frac{B_2 m_1'}{\rho M_{\rm s}}, \\
    -\omega^2 A_2 &= c_S^2 (A_2''-k^2 A_2) + \frac{B_2 m_2'}{\rho M_{\rm s}}, \\
   -\omega^2 A_3 &= c_L^2  A_3'' -  c_S^2 k^2 A_3 + (c_L^2-c_S^2) i k A_1' + \frac{i k B_2 m_1}{\rho M_{\rm s}}, \\
    \label{eq:mag_1}
    i \omega m_1 &= \gamma \mu_0  \left[(H_0-M_{\rm s}+\ell^2 M_{\rm s}k^2)m_2 -\ell^2 M_{\rm s} m_2'' + \frac{B_2 A_2'}{\mu_0}\right], \\
    \label{eq:mag_2}
    i \omega m_2 &= \gamma \mu_0 \left[(-H_0+M_{\rm s}-\ell^2 M_{\rm s} k^2)m_1 +\ell^2 M_{\rm s} m_1''-i k M_{\rm s} g  - \frac{B_2 A_1'}{\mu_0} + \frac{i k B_2 A_3}{\mu_0}\right],
\end{align}\end{subequations}    \end{widetext}
where $g$ satisfies \eqref{eq:g_equation}.
Equations \eqref{eq:full_coupled_system} and \eqref{eq:g_equation} are a system of six ordinary differential equations for six variables in a magnetoelastic layer. Note that the first coupling coefficient $B_1$ plays no role in the linearized system. For a purely elastic layer, $B_2=0$, and the system reduces to the three elastic equations.

\subsection{Boundary conditions}
We must also incorporate boundary conditions to close the above
system. For a purely elastic layer, we have three second-order
equations, requiring six boundary conditions. These come from
continuity of the three stresses $\sigma_{j3}$ \eqref{eq:el_gen_bc}
and three displacements $A_j$, $j \in \{1,2,3\}$ at a boundary. If the
boundary is with vacuum, we instead require that $\sigma_{j3}=0$ at
those boundaries. The magnetoelastic stresses are calculated as
$\sigma = \delta (E_c+E_{\rm el})/\delta \varepsilon$, yielding
\begin{subequations}
  \label{eq:17}
  \begin{align}
    \label{eq:18}
    \sigma_{13} &= \mu(A_1'+ikA_3) + \frac{B_2}{M_{\rm s}} m_1, \\
    \label{eq:19}
    \sigma_{23} &= \mu A_2' + \frac{B_2}{M_{\rm s}} m_2, \\
    \label{eq:20}
    \sigma_{33} &= \lambda ik A_1 + (\lambda+2\mu)A_3' .
  \end{align}
\end{subequations}
For a magnetic layer, we include the Neumann conditions for
magnetization elements \eqref{eq:mag_gen_bc}
$\partial m_j/\partial x_3 = 0$, $j \in {1,2}$, along with the Robin
conditions for the dipole field \eqref{eq:g_equation}.

Since eigenfunctions are multiplied by a complex exponential, the mode
profile is the real part of
$\begin{bmatrix}A_1&A_2&A_3&m_1&m_2\end{bmatrix} e^{i(kx_1-\omega
  t)}$.  Based on the symmetry properties of equations
\eqref{eq:full_coupled_system} and the associated boundary conditions,
eigenfunctions can be normalized so that
\begin{equation}
  \label{eq:16}
  \mathrm{Re} \begin{bmatrix} A_1\\m_1 \end{bmatrix} = 0 \quad
  \mathrm{and} \quad
  \mathrm{Im} \begin{bmatrix} A_2\\A_3\\m_2 \end{bmatrix} = 0.
\end{equation}
Consequently, modes oscillate between populating the longitudinal
components $A_1$, $m_1$ and then the transverse components $A_2$,
$A_3$, and $m_2$, which are $\pi/2$ out of phase.  Due to the property
\eqref{eq:16}, taking the imaginary part of the boundary condition
\eqref{eq:18} implies that $A_1'$ can be discontinuous across the
interface between a magnet and non-magnet.

\section{Analytical calculations}
\label{sec:analysis}


\subsection{Bulk magnetoelastic waves}
\begin{figure}
    \centering
    \includegraphics[scale=.25]{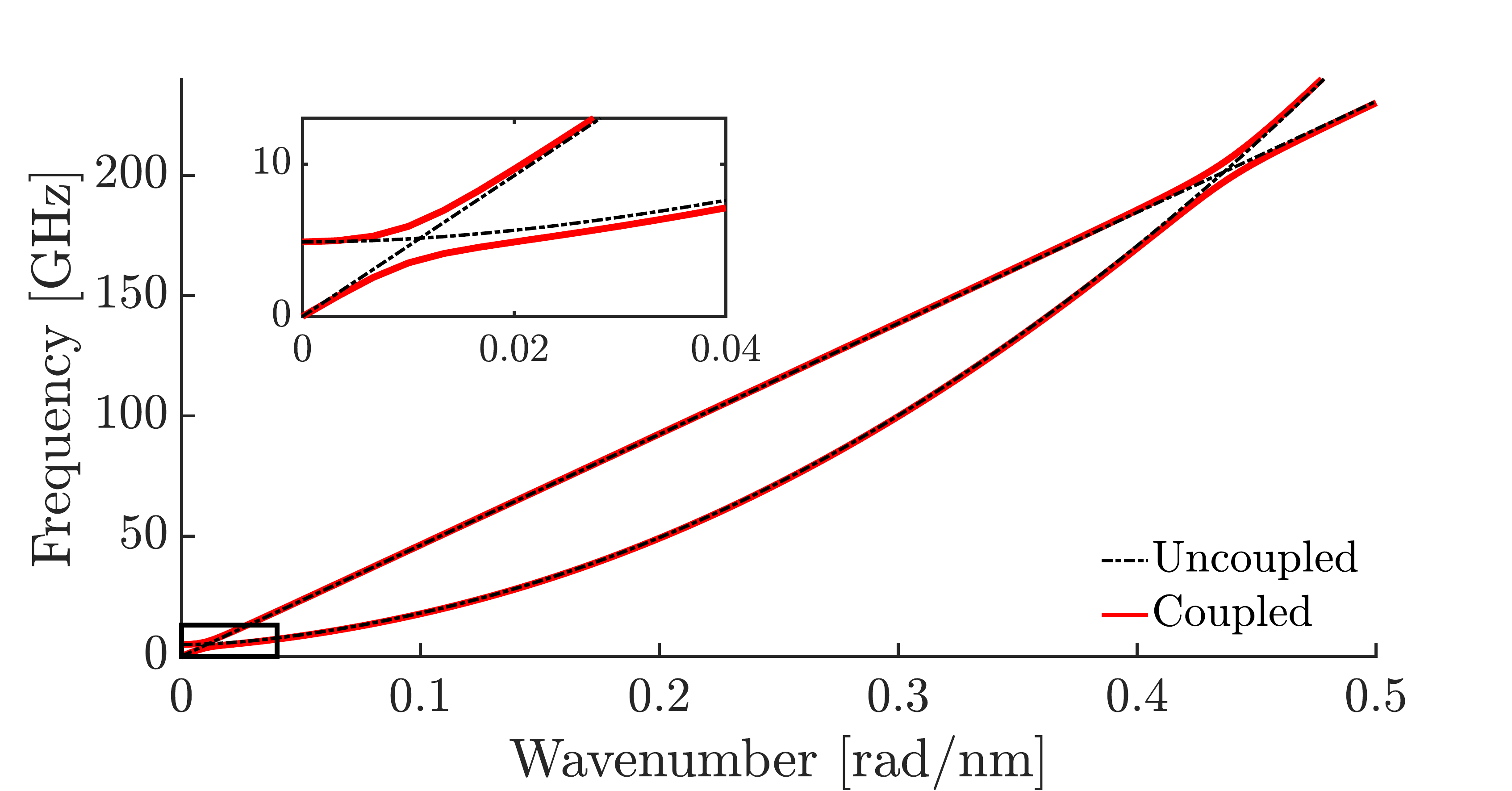}
    \caption{Coupled (solid) and uncoupled (dashed-dotted) dispersion
      curves for bulk nickel with an applied field perpendicular to
      the direction of wave propagation. The uncoupled curves have two
      interactions, where for coupled curves anticrossings appear. The
      lower-frequency anticrossing is shown zoomed-in in the
      inset. The applied field has magnitude $\mu_0 H_0=0.65$ T.  The
      coupling coefficient $B_2$ here is chosen to be unphysically
      large ($B_2 = 20$ MJ/m$^3$) in order to visually emphasize
      anticrossing behavior; all other material constants are as in
      Table~\ref{table:speeds}.}
    \label{fig:nickel_regimes}
\end{figure}
The dispersion relations for magnetoelastic waves in a bulk material
with $d \to \infty$ can be calculated directly from
\eqref{eq:full_coupled_system}. In this case, all derivatives with
respect to $x_3$ are assumed to be negligibly small, and boundary
effects are ignored. The resulting sixth-order homogeneous linear
system for $A_i$, $m_j$, and $g$ requires a nonzero determinant in
order for there to be a nonzero solution. This relation then yields a
sextic polynomial equation for $\omega$.
 
First, we consider uncoupled waves $(B_2 = 0)$. For the three
purely elastic waves, the dispersion then is
\begin{equation}
  \label{eq:bulk_el}
  \omega_{\rm el} = c k,
\end{equation}
where $c$ is one of the speeds of sound \eqref{eq:speeds of
  sound}. For $A_1$, the wave is longitudinal and travels with speed
$c_L$, while for $A_2$ and $A_3$ the wave is transverse (shear) and
travels with speed $c_S$. The dispersion for magnetic waves in a bulk
material with a perpendicular applied field is
\begin{equation}
    \label{eq:bulk_mag}
    \omega_{\rm m} = \pm\left[(\omega_{\rm H}- \omega_{\rm M} + \beta k^2)(\omega_{\rm H} + \beta k^2 ) \right]^{1/2},
\end{equation}
where 
\begin{equation}
  \label{eq:2}
  \omega_{\rm H} = \gamma \mu_0 H_0, \quad \omega_{\rm M} = \gamma
  \mu_0 M_{\rm s}, \quad \beta = \gamma \mu_0 \ell^2 M_{\rm s}.  
\end{equation}
Note that \eqref{eq:bulk_mag} differs slightly from what is typically
presented (cf. \cite{gurevich_1996}), since the $d \to \infty$ limit
of our equations \eqref{eq:full_coupled_system} retains a $-M_{\rm s}$
term due to the demagnetizing effect of the material boundaries.

Next, we consider the dispersion of \eqref{eq:full_coupled_system}
with nonzero magnetoelastic coupling $(B_2 \neq 0)$. In the bulk
limit, $A_1$ and $A_2$ separate from the magnetic terms and the
corresponding waves travel with the uncoupled speeds $c_L$ and $c_S$,
respectively. The remaining elastic component $A_3$ and the magnetic
components $m_1$ and $m_2$ form coupled magnetoelastic waves subject
to the quartic dispersion relation
\begin{equation}
  (\omega^2-\omega_{\rm m}^2)(\omega^2-\omega_{\rm el}^2) -
  \frac{\gamma B_2^2}{\rho M_{\rm s}}(\omega_{\rm H}-\omega_{\rm
    M}+\beta k^2) k^2 = 0,
\end{equation}
where $\omega_{\rm el} = c_S k$. In Fig.~\ref{fig:nickel_regimes}, we
show the positive, uncoupled and coupled magnetoelastic dispersion
curves for $A_3$ and magnetism in a bulk material subject to a
perpendicular applied field.

The coupled and uncoupled bulk dispersion curves present some general
principles regarding magnetoelastic wave interactions. As evident from
Fig.~\ref{fig:nickel_regimes}, magnetoelastic interactions are most
significant near the intersections of the uncoupled magnetic and
elastic dispersion curves where anticrossings appear. Instead of
intersecting, the dispersion curves bend away from one another,
leaving a gap. The gap width between the curves is determined by the
strength of the magnetoelastic coupling $B_2$
\cite{gurevich_1996,vanderveken_confined_2021}. Far away from the
anticrossing, the dispersion curves follow the uncoupled magnetic or
elastic curves and are known as quasi-magnetic and quasi-elastic,
respectively. Near the anticrossings, the quasi-elastic and
quasi-magnetic dispersion curves bend away from each other and switch
roles. Thus, a wave with energy primarily concentrated in elastic
oscillations transfers that energy into magnetic waves when the
frequency is through an anticrossing, while the corresponding magnetic
wave transfers its energy into elastic vibrations.  Two examples of
anticrossings are shown in Fig.~\ref{fig:nickel_regimes}, one at lower
frequency and one at higher frequency.

More generally, uncoupled magnetic dispersion curves
\eqref{eq:bulk_mag} are approximately parabolic, while uncoupled
elastic curves \eqref{eq:bulk_el} are approximately linear. These
curves have two intersections, so there are two regimes at which
magnetoelastic interactions occur. The first intersection often occurs
in the single or tens of gigahertz regime, where dipole effects are
comparable or stronger than exchange. Due to the longer wavelengths,
these interactions can often be studied in finite-thickness samples
analytically using a thin film assumption
\cite{vanderveken_confined_2021}. These lower-frequency interactions
have been studied experimentally; see, for example
\cite{weiler_2011_resonance,kamra_2015_elastic_excitation,kim_magnetization_2017,janusonis_simultaneous_2016,holanda_conversion_2018,an_momentum_2020,godejohann_magnon_2020,zhang_high-frequency_2020}. The
second dispersion curve intersection occurs in a much higher frequency
regime, where exchange effects dominate. In this work, we will focus
on the higher-frequency interactions, which are not as well-studied
but are important for understanding the recovery of magnetic order
post ultrafast demagnetization \cite{turenne_nonequilibrium_2022}.

Even when layering is incorporated, we still expect to see
magnetoelastic interactions in the $\omega$-$k$ region near where the
bulk shear and magnetic dispersion curves intersect, i.e.
$\omega_{\rm m} \approx \omega_{\rm el}$.  For YIG and Ni, the
higher-frequency intersections between the shear elastic and magnetic
waves are around 100-200 GHz, and so are well within the current
experimentally accessible regime. The longitudinal speeds of sound are
twice as large (see Tab.~\ref{table:speeds}).  Consequently,
longitudinal-magnetic waves are expected to occur at twice the
frequency range. The separation of $A_1$ and $A_2$ from the other wave
components in the bulk limit indicates that the most significant
elastic-magnetic interactions will involve the $A_3$ shear elastic
component and $m_1$, $m_2$.

We also note that the higher-frequency magnetic and elastic
intersections in Fig.~\ref{fig:nickel_regimes} for Ni occur at
wavelengths $2\pi /k\approx 15$ nm. This is approximately twice the
exchange length $\ell$ for nickel (see
Tab.~\ref{table:speeds}). Consequently, we expect the exchange energy
to be more significant than the long-range dipole energy.
Nevertheless, dipole effects must still be considered. A similar
argument holds for YIG which has an exchange length $\ell\approx$ 17
nm, and its magnetoelastic intersections occur at a similar
wavelength.

In summary, analysis of the bulk dispersion curves reveals that
magnetoelastic anticrossings between magnetic and shear waves are
possible in the extremely high frequency regimes of experimental
interest and exchange effects will be more significant than long-range
dipole effects in this regime.

\subsection{Single layer asymptotic calculation}
\label{sec:mag_asymp}

When finite-thickness and boundary effects are included, a direct
analytical calculation of the dispersion from the eigenvalue problem
\eqref{eq:g_equation}, \eqref{eq:full_coupled_system} is challenging.
In contrast to the bulk limit, a finite-thickness film in the
continuum approximation has an infinite number of dispersion curves,
corresponding to higher-order quasi-elastic and quasi-magnetic modes.
Of course, the actual number is limited by the crystal lattice
spacing. In this section, we utilize an asymptotic calculation that is
presented in Appendix \ref{sec:asympt-calc-exch} to show that, unlike
in the bulk limit, not all magnetic-elastic curve intersections yield
anticrossing behavior.

Motivated by the analysis in the previous subsection, we consider a
simplified scenario where displacements in the $A_1$ and $A_2$
direction are ignored.  We also neglect dipole effects, i.e., we set
$g \equiv 0$. These assumptions greatly simplify the eigenvalue
problem \eqref{eq:g_equation} and \eqref{eq:full_coupled_system}, yet
they reveal significant insight into the problem.  The uncoupled case
($B_2 = 0$) gives rise to the discrete, infinite family of elastic
$\omega_{{\rm el},n}$ and magnetic $\omega_{{\rm m},j}$ dispersion
branches
\begin{subequations}
  \label{eq:layer_uncoupled_dispersion}
  \begin{align}
    \label{eq:layer_uncoupled_elastic}
    \omega_{{\rm el},n} &= \sqrt{(c_S k)^2 + (c_L \xi_n)^2}, \quad n
                          = 0,1,\ldots \\
    \label{eq:layer_uncoupled_magnetic}
    \omega_{{\rm m},j} &= \omega_{\rm H} - \omega_{\rm M} + \beta (k^2
                         + \xi_j^2 ), \quad j = 0,1,\ldots ,
  \end{align}
\end{subequations}
where $\xi_n = n\pi/d$ is the discrete vertical wavenumber.  The
corresponding vertical mode profiles are cosines
\begin{subequations}
  \label{eq:modes}
  \begin{align}
    \label{eq:elastic_mode}
    A_{3,n}(x_3) = a_{{\rm el},n}\cos(\xi_n x_3), \quad n = 0, 1,
    \ldots ,\\
    \label{eq:magnetic_mode}
    m_{2,j}(x_3) = a_{{\rm m},j}\cos(\xi_j x_3), \quad j = 0, 1,
    \ldots .
  \end{align}
\end{subequations}

Magnetic materials exhibit weak magnetoelastic coupling $B_2$.  We
find in App.~\ref{sec:asympt-calc-exch} that this can be quantified by
the smallness of the nondimensional parameter
\begin{equation}
  \label{eq:epsilon}
  \epsilon = B_2 \frac{\gamma \ell}{c_S}\sqrt{\frac{\mu_0}{\rho}} \ll
  1 .
\end{equation}
For example, $\epsilon \approx 0.02$ for both YIG and Ni.  An
expansion of the frequencies and mode profiles around
\eqref{eq:layer_uncoupled_dispersion} and \eqref{eq:modes} for small
$\epsilon$ determines the following resonance condition
\begin{equation}
  \label{eq:resonance}
  \omega_{{\rm el},n}(k) \approx \omega_{{\rm m},j}(k), \quad n = j,
\end{equation}
necessary for magnetoelastic frequency corrections that are
proportional to $\epsilon$.  Otherwise, for nonresonant conditions,
the frequency corrections are significantly smaller (proportional to
$\epsilon^2$).  Equation \eqref{eq:resonance} represents two resonance
requirements: i) the wavenumber $k$ must be close to $k_*$, an
intersection of the uncoupled elastic and magnetic dispersion curves,
ii) the mode numbers for the elastic and magnetic waves must be the
same $n = j$.  The frequency corrections are provided in
App.~\ref{sec:asympt-calc-exch} (see eqs.~\eqref{eq:as_expand},
\eqref{eq:as_comp}) and describe an anticrossing.

Additionally, the resonance condition \eqref{eq:resonance} leads to a
wavenumber dependent relative scaling between the elastic and
$a_{{\rm el},n}$ and magnetic $a_{{\rm m},n}$ mode amplitudes (see
eq.~\eqref{eq:rel_sizes}).  Combining this scaling and the frequency
corrections enables one to describe how the frequency transitions from
a quasi-elastic or quasi-magnetic branch to the other.  The evaluation
of the frequency corrections at the point of intersection $(k_*,\omega_*)$
\begin{equation}
  \label{eq:4}
  \omega_* = \omega_{{\rm el},n}(k_*) = \omega_{{\rm m},n}(k_*)
\end{equation}
of the uncoupled dispersion branches
\eqref{eq:layer_uncoupled_dispersion} determines the anticrossing
frequency gap width
\begin{equation}
  \label{eq:3}
  \Delta_{\rm gap} = \epsilon \frac{c_S^2 k_*}{\sqrt{\beta \omega_*}} .
\end{equation}
The gap width is proportional to $B_2$, as in the bulk and thin film
limits \cite{gurevich_1996,vanderveken_confined_2021}.  In addition,
the gap width is proportional to the wavenumber $k_*$ and inversely
proportional to the square root of the frequency $\omega_*$ at the
intersection \eqref{eq:4}.  Consequently, larger wavenumber and
smaller frequency resonances give rise to larger anticrossing gaps.

\begin{figure}
  \centering
  \includegraphics[scale=.25]{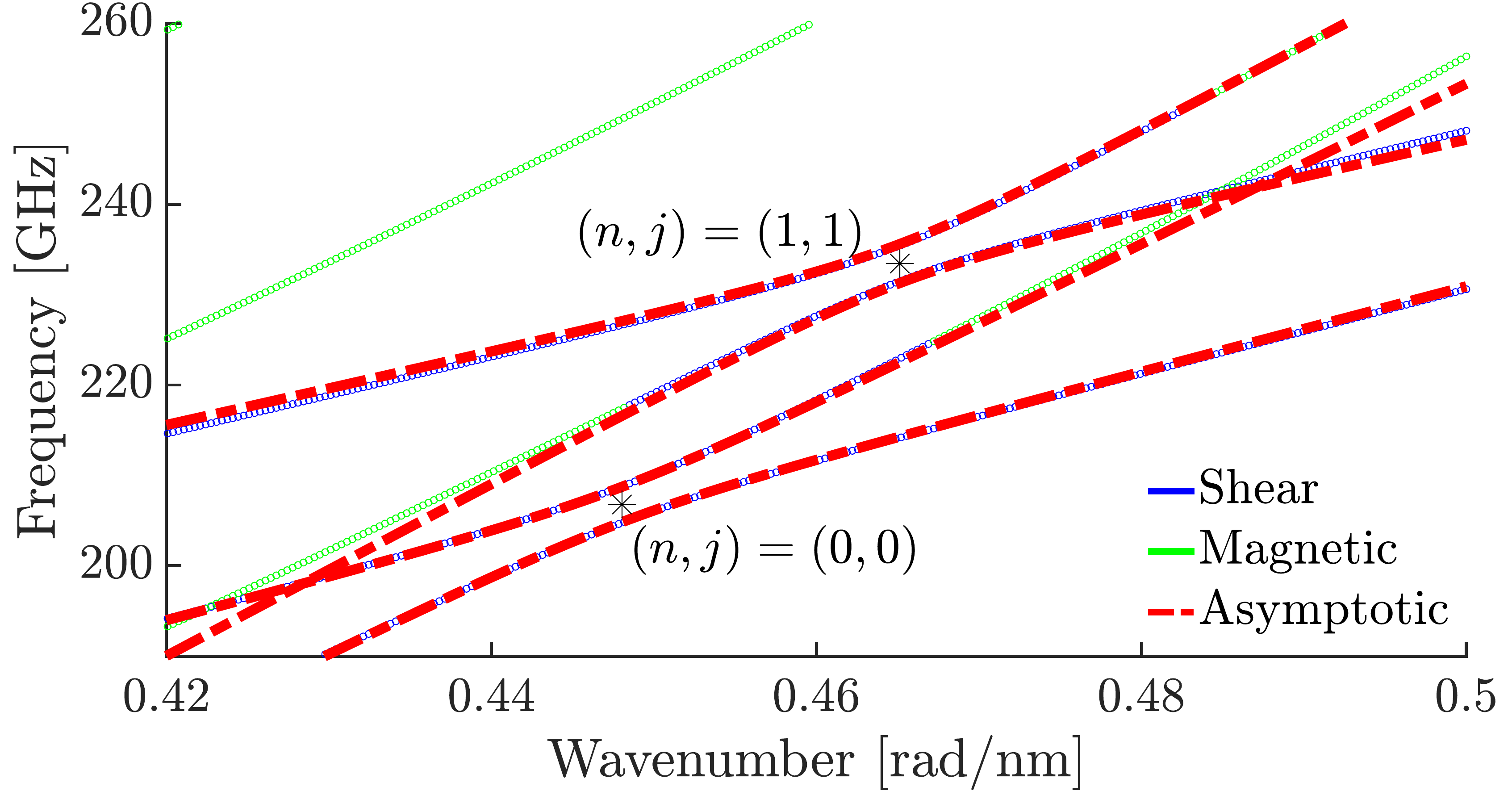}
  \includegraphics[scale=.25]{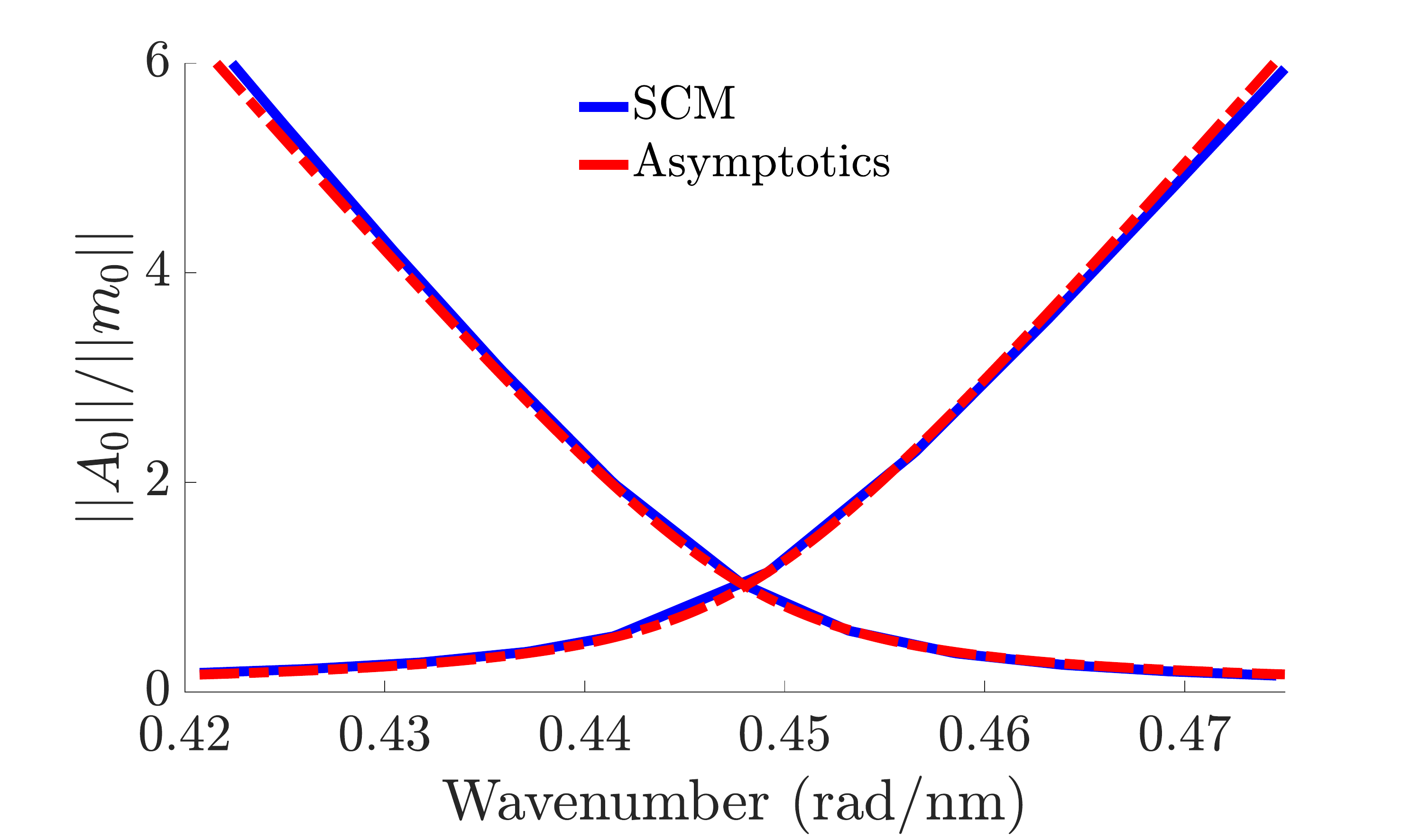}
  \caption{Top: comparison of the asymptotic prediction
    \eqref{eq:as_comp} and \eqref{eq:as_expand} (dashed-dotted) with
    dispersion curves calculated by SCM (solid) for the two lowest
    modes $n = 0,1$ and $j = 0,1$. The material thickness is $d=30$
    nm, and the applied field is $\mu_0 H_0=0.2$ T. Intersection
    points are labeled. Bottom: relative wave magnitudes for the
    $(n,j)=(0,0)$ resonant intersection shown on the left plot as
    determined from numerics (solid) and asymptotics in
    \eqref{eq:rel_sizes} (dashed-dotted).}
  \label{fig:as_comp}
\end{figure}
We compare asymptotic results with the SCM computation, described in
the next Section, of the simplified system considered here
analytically (eq.~\eqref{eq:54} with elastic components $A_1$, $A_2$
and dipole field $g$ neglected) for a single layer of YIG.  The top
panel of Fig.~\ref{fig:as_comp} shows that, in the vicinity of
$n = 0,1$ resonant interactions, the agreement between the asymptotic
prediction (dash-dotted) and the SCM calculated dispersion curves
(solid) is excellent.

On the bottom panel of Fig.~\ref{fig:as_comp}, we show the relative
magnitudes of the nondimensionalized elastic and magnetic waves for
the $(n,j)=(0,0)$ intersection depicted in the top panel. The
numerical calculations (solid) for the norm ratio agrees very well
with the asymptotic predictions (dash-dotted).

\section{Spectral Collocation Method}
\label{sec:scm}
The previous section demonstrated a marked increase in dispersion
complexity due to the incorporation of boundary effects for a single,
finite-thickness magnetoelastic layer. In this Section, we introduce a
spectral collocation method (SCM) for determining the dispersion
relations of multiple layers. Unlike the above asymptotic calculation,
this method studies the full system \eqref{eq:full_coupled_system},
including the dipole field \eqref{eq:g_equation}.

\subsection{Method overview}

The SCM approach we develop is adapted from one utilized for purely
elastic waves
\cite{adamou_2004,quintanilla_2015,quintanilla_2015_spectral}. Consider
$N$ points $x^{(j)}$ on the domain $[-1,1]$ given by the formula
\begin{equation}
    \label{eq:cheby_points}
    x^{(j)} = \cos{\frac{(j-1)\pi }{N-1}}, \qquad j = 1,\dots,N.
\end{equation}
These are known as Chebyshev points (or Gauss-Chebyshev-Lobatto
points) \cite{trefethen_matlab}. They arise as the roots of the
$N^{\rm th}$ degree Chebyshev polynomial. Their distribution over
$[-1,1]$ is nonuniform and concentrated near the endpoints of the
interval. More details about Chebyshev points and Chebyshev polynomial
interpolation can be found in \cite{trefethen_matlab,cheby_book}.

For each layer, the six functions $A_1(x_3)$, $A_2(x_3)$, $A_3(x_3)$,
$m_1(x_3)$, $m_2(x_3)$, and the dipole field $g(x_3)$ are interpolated
at Chebyshev points \eqref{eq:cheby_points} into vectors of length
$N$,
\begin{equation}
    \bm{A_1}=\begin{bmatrix}
    A_1\left(x^{(1)}\right) &
    \dots &
    A_1\left(x^{(N)}\right)
    \end{bmatrix}^{\rm T},
 \dots
\end{equation}
The key to the efficiency of the SCM method is that, for smooth
functions, a discretization at Chebyshev nodes \eqref{eq:cheby_points}
converges faster than any negative power of $N$, i.e. the convergence
is spectral. In fact, one can show that the Chebyshev points are
nearly optimal for minimizing the uniform norm of the interpolation
error \cite{trefethen_matlab}.

Each derivative with respect to $x_3$ of the discretized functions can
be approximated using a Chebyshev differentiation matrix $D_N$. These
matrices can be efficiently and stably generated using the \emph{cheb}
function in MATLAB \cite{trefethen_matlab}. The differentiation matrix
$D_N$ generated using the MATLAB function \emph{cheb} must be scaled
by $2/(b-a)$ to account for a general domain $[a,b]$. Concatenating
the matrices and vectors $\bm{A_1},\dots,$ yields a quadratic
polynomial eigenvalue problem of the form
$(\mathcal{A}_2 \omega^2+\mathcal{A}_1 \omega + \mathcal{A}_0)
\bm{\phi} = 0$, where $A_j$, $j=0,1,2$ are dense $6N \times 6N$
matrices for a single magnetic layer.

Next, the boundary conditions are incorporated by replacing the rows
representing the boundaries for each of the vector equations. The
boundary rows occur at rows $1,N,N+1,2N,\dots, 6N$ and are discretized
with Chebyshev differentiation matrices as above. Then these equations
are inserted into the $1,N,N+1,2N,\dots,6N$ rows of the polynomial
eigenvalue problem. Since our boundary conditions are time-independent
and therefore frequency independent, the boundary effects are all
incorporated into $\mathcal{A}_0$, and the corresponding
$1,N,N+1,\dots$ rows of $\mathcal{A}_2$ and $\mathcal{A}_1$ are
replaced with zeros.

Polynomial eigenvalue problems can be solved directly using the
built-in MATLAB \emph{polyeig} command. The frequency polynomial is
second degree, so in principle we will find $12N$ modes for a coupled
magnetoelastic layer. An increase in $N$ leads to an increase in the
number of approximated eigenvalues, with new, larger eigenvalues
appearing while the smaller eigenvalues exhibit improved accuracy. Due
to spectral convergence, $N$ need not be very high before accurate
results are obtained for the smaller eigenvalues.

In the polynomial eigenvalue problem, the linear in $\omega$ terms
arise from the Landau-Lifshitz equation \eqref{eq:landau_lifshitz},
which is a first order in time system of differential equations. The
quadratic $\omega^2$ terms arise from the Navier equation
\eqref{eq:navier}, which is second order in time. Previous spectral
collocation methods applied to purely elastic problems
(e.g.~\cite{quintanilla_2015_spectral}) only required the solution of
an eigenvalue problem for $\omega^2$, not a polynomial eigenvalue
problem, since the square root and splitting into two $\pm$ branches
can be computed afterward.

The extension of the above method to multiple layers is
straightforward. We outline it for a magnetoelastic material layered
on a purely elastic substrate. The substrate has three governing
equations for the displacements $\bar A_j$, $j = 1,2,3$ (we denote
displacements in the substrate by an overbar), for a total of nine
equations. The discretized field matrix for these equations
($3N \times 3N$) is joined to the discretized field matrix above to
form a matrix with $(9N)^2$ elements. The differentiation matrix in
each layer $D_N$ must be suitably scaled to account for the
thicknesses of the two layers. Once again, the appropriate boundary
conditions replace rows $1,N,N+1,\dots,9N$. The larger quadratic
eigenvalue problem is solved to obtain the solution.

The generated matrices in the polynomial eigenvalue problem can be
ill-conditioned, so the problem is numerically sensitive. In order to
ensure sufficient numerical precision, we utilize a multi-precision
MATLAB toolbox to perform calculations in quadruple precision. Higher
precisions than quadruple did not appear to increase the accuracy, and
the toolbox is calibrated to perform optimally in quadruple precision
\cite{mct2015}.  Moreover, we found that many of our calculations
yield similarly accurate results even for double precision
computations.

One advantage of the SCM method is the direct computation of the
dipole field $g$. Accounting for dipole effects is one of the most
significant challenges when studying magnetic waves. By including $g$
as a discretized function in the polynomial eigenvalue problem, we
spectrally converge to the correct dipole field with little additional
computational cost.

Another benefit of the SCM method is that the eigenvector determined
from the polynomial eigenvalue problem is a composite vector
containing the discretized vertical mode profiles
$\bm{A}_1,\bm{A}_2,\dots, \bm{m}_1, \dots$ at the Chebyshev points
\eqref{eq:cheby_points}. We utilize this fact for two purposes. First,
we can compare the relative 2-norms of these vectors to classify the
mode type corresponding to a particular dispersion curve. For example,
if $||\bm{A_1}||\gg ||\bm{A_2}||$, the displacement is predominately
in the $x_1$-direction, and the wave can be considered
longitudinal. Using the scaling between magnetism and elasticity
determined from the nondimensionalization \eqref{eq:nondim}, we can
compare $||\bm{A_i}||/A_*$ and $||\bm{m}_j||/M_{\rm s}$ to determine
whether a mode is predominately elastic or magnetic. Due to the
polarization of magnetic waves, all calculations involving the vectors
$\bm{m_1}$ and $\bm{m_2}$ instead utilize the 2-norm of the
magnetization vector, $\bm{m} = \sqrt{|\bm{m}_1|^2+|\bm{m}_2|^2}$. In
the below figures displaying results of the SCM method, each
dispersion curve is labeled by color to identify the eigenvector's
component that is largest, with the appropriate scalings given by
\eqref{eq:nondim}. This classification method is simple, and the below
figures will indicate its effectiveness.

The discretized eigenvector can be used to determine the vertical mode
profile by using the Chebyshev interpolating polynomial corresponding
to the discretization. To ensure the property \eqref{eq:16}, we first
apply a complex phase shift so that the vertical shear profile
evaluated at the top of the upper layer is positive $A_3(d) > 0$.
Then, we present the real parts of $A_2$, $A_3$, and $m_2$ together
and, separately, the $\pi/2$ out of phase imaginary parts of $A_1$ and
$m_1$.  In order to compare profiles of elastic and magnetic
components, they are first nondimensionalized according to
(cf.~eq.~\eqref{eq:60})
\begin{equation*}
  A_i/A_*, \quad i = 1,2,3, \quad m_j/M_{\rm s}, \quad j = 1,2,
\end{equation*}
and then normalized by the maximum amplitude. Wave profiles are
presented below in Figs.~\ref{fig:one_layer_res},
\ref{fig:wave_struc}, and \ref{fig:low_freq_anticrossing}.

A validation study is performed in Appendix \ref{sec:validation-scm}.
We observe rapid convergence of the frequencies requiring only a
modest number of Chebyshev discretization points
$N \in \{10,\ldots,24 \}$.  Furthermore, we directly compare the
uncoupled $B_2 = 0$ elastic dispersion computed using SCM for a double
layer with a published elastic dispersion solver, finding quantitative
agreement.

\section{Results}
\label{sec:results}

In this section, we review some selected results of SCM calculations
in order to highlight its utility. 
In all figures in this section, curves with mode profiles whose
2-norms are dominated by magnetism are green, elastic waves in a
magnetic material are blue, and elastic waves in a nonmagnetic
material are red. In addition, variables corresponding to a
nonmagnetic material are denoted by a bar, e.g., $\overline{A}_1$.

\subsection{Single layer}
\subsubsection{Dispersion map}

\begin{figure}
  \centering \includegraphics[trim={0 3cm 0
    0},clip,scale=.25]{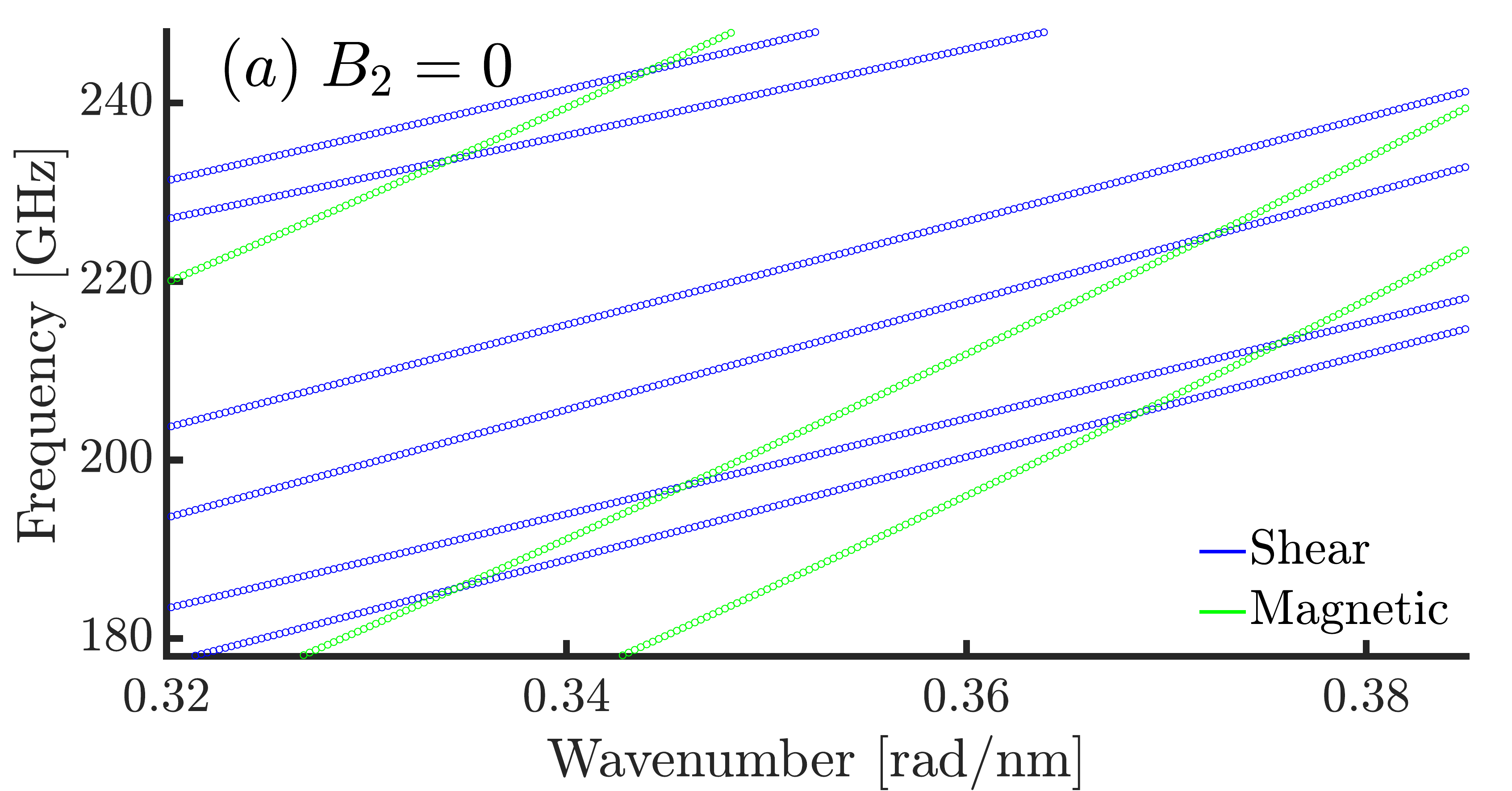}
  \includegraphics[scale=.25]{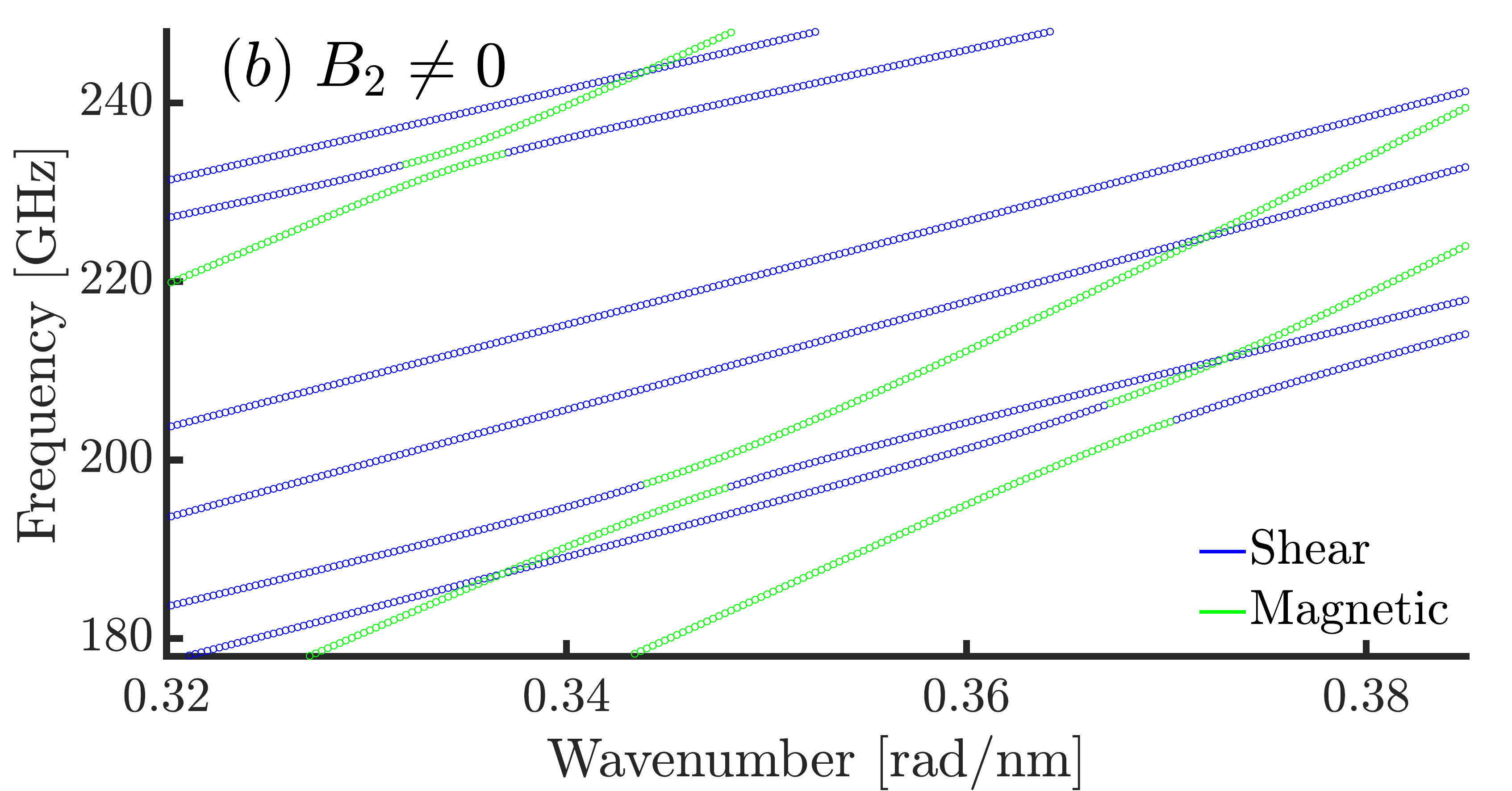}
  \caption{Dispersion curves for one layer of YIG with thickness 30 nm
    and $N=16$ with coupling constant $B_2 =0$ for $(a)$ and the
    physical value of $B_2$ in Tab.~\ref{table:speeds} for
    $(b)$. Three anticrossings and multiple ordinary crossings are
    visible.}
  \label{fig:single_layer_compare}
\end{figure}

We first examine the effect of magnetoelastic coupling on a single
layer. Figure~\ref{fig:single_layer_compare} shows SCM calculations
with $N=16$ for a single layer of YIG with thickness 30 nm and an
applied field of 0.25 T. Panel $(a)$ shows the uncoupled result when
$B_2 = 0$, resulting in purely elastic or magnetic modes. The
dispersion curves cross without any interaction.

In contrast, panel $(b)$ has a nonzero coupling constant $B_2$ set to
the physical value for YIG given in Tab.~\ref{table:speeds}. Away from
the intersections, the dispersion curves are nearly identical between
panels $(a)$ and $(b)$. Near some intersections, however,
anticrossings appear, as expected. Multiple simple crossings are also
displayed.  One can readily observe that the anticrossing gap width
decreases with increasing frequency and lower wavenumber.  These
findings are consistent with the simplified analysis presented in
Sec.~\ref{sec:analysis}.

\subsubsection{Resonant and nonresonant interactions}
\begin{figure*}
  \centering
  \includegraphics[scale=.25]{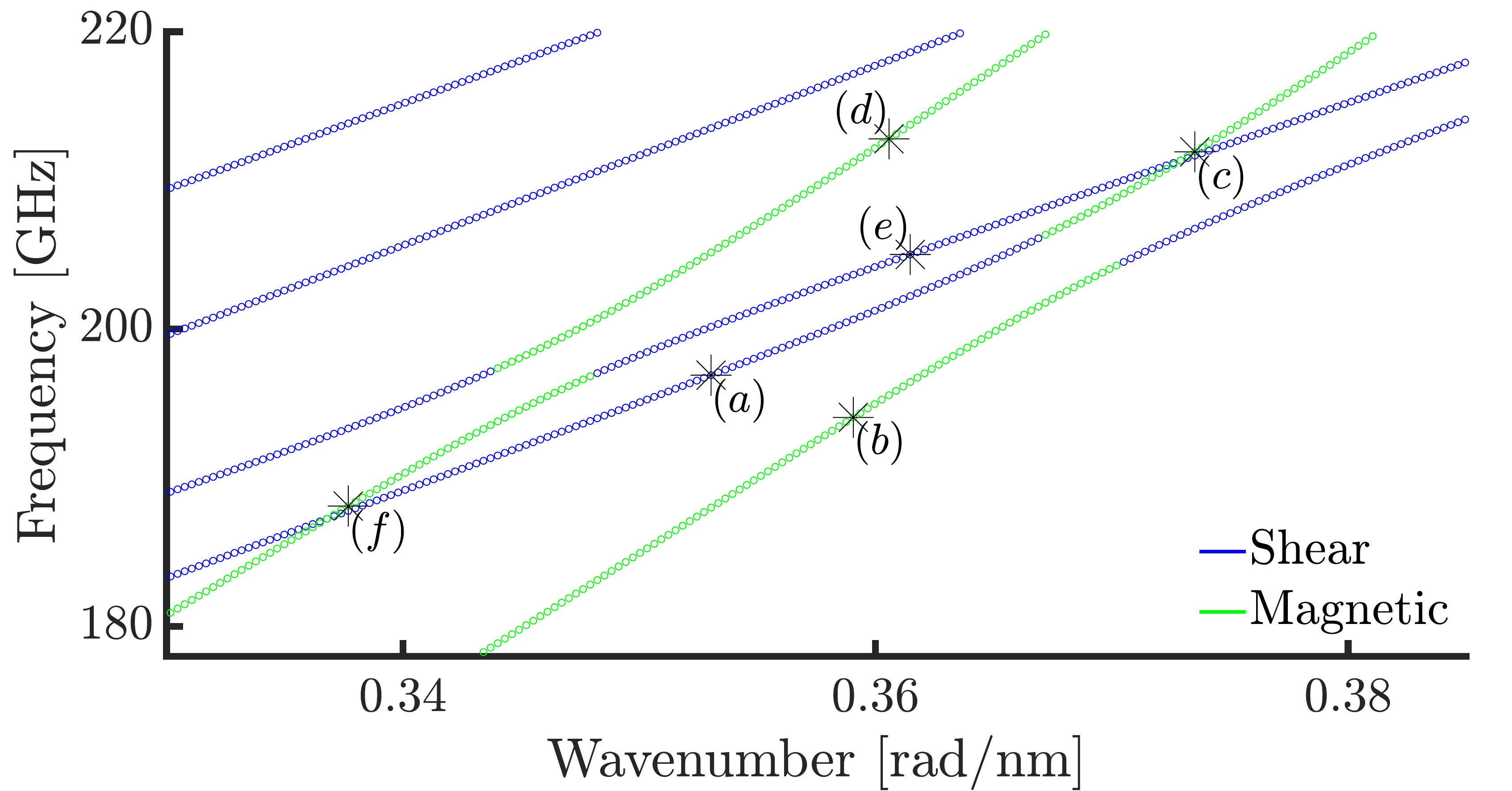}\\
  \includegraphics[scale=.25]{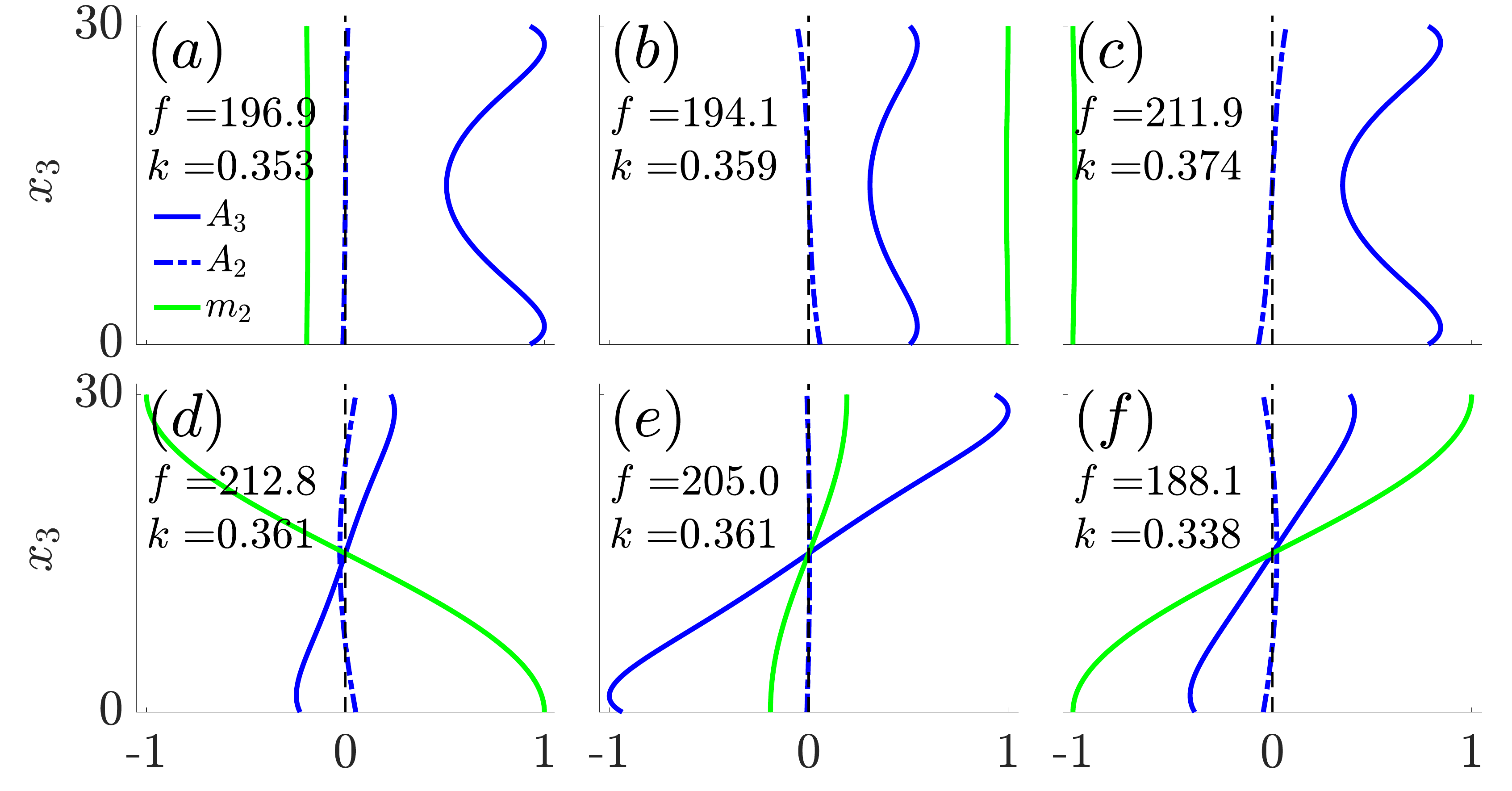}
  \includegraphics[scale=.25]{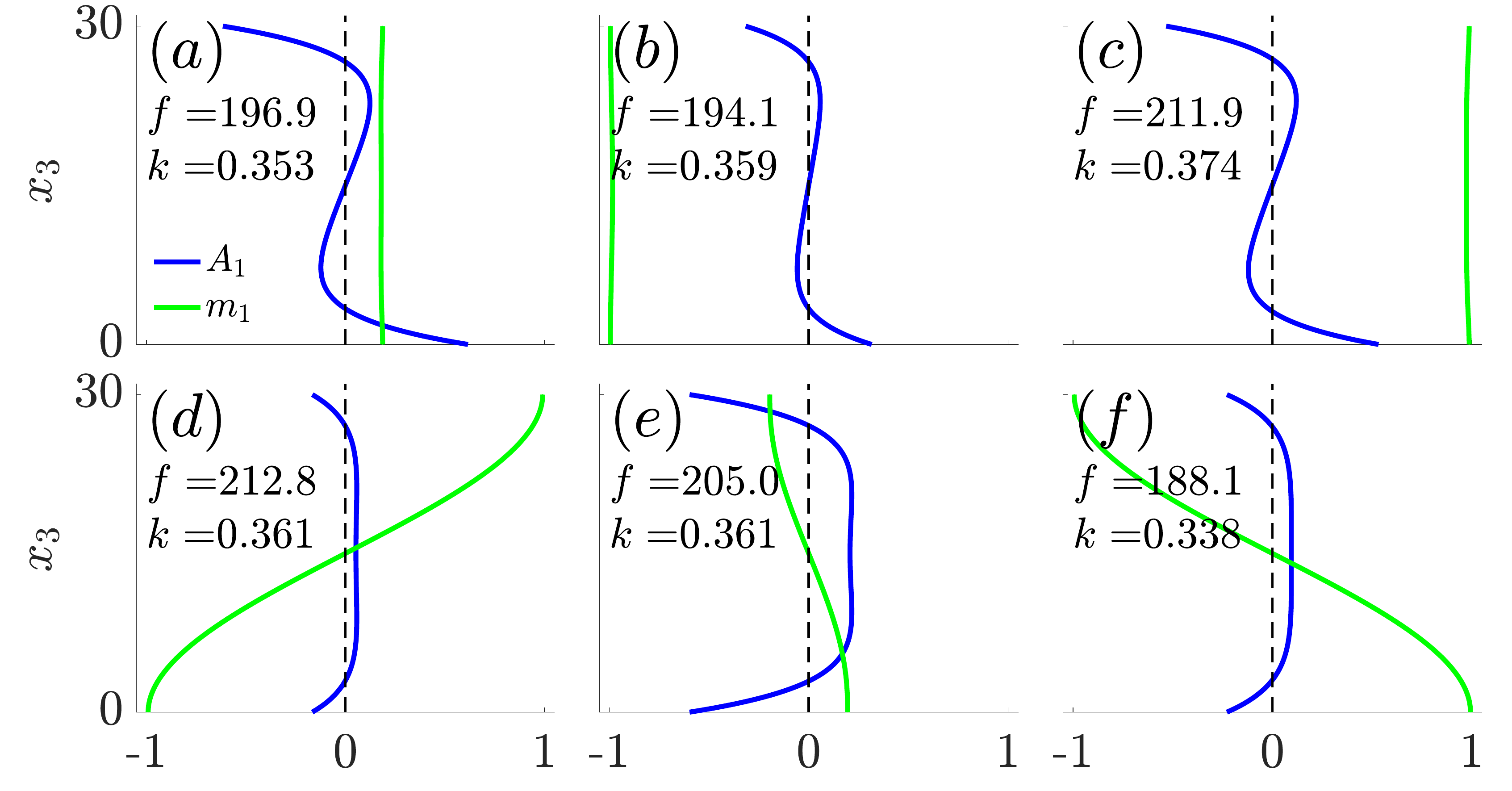}
  \caption{Top: zoomed in anticrossings from
    Fig.~\ref{fig:single_layer_compare}$(b)$. Bottom: mode profiles
    corresponding to the dispersion points $(a)$--$(f)$ with
    $(A_3,A_2,m_2)$ (Left) and the $\pi/2$ phase advanced profiles
    $(A_1,m_1)$ (Right).  Points $(a)-(c)$ correspond to an
    anticrossing in the zeroth order mode of $A_3$ and $(m_1,m_2)$, so
    the wave profiles do not exhibit zeros. Points $(d)-(f)$
    correspond to the first order mode in $A_3$ and $(m_1,m_2)$, so
    their profiles have a single zero. Figures $(c)$ and $(f)$ are the
    profiles at nonresonant crossings; the nonresonant correction is
    higher order and not visible in this figure.}
    \label{fig:one_layer_res}
\end{figure*}
We next examine resonant and nonresonant interactions in the single
layer of YIG. In the top panel of Fig.~\ref{fig:one_layer_res}, we
show a zoomed in view of the two lowest-order anticrossings from
Fig.~\ref{fig:single_layer_compare}$(b)$. The wave profiles at the
labeled points are shown in the bottom panels. For the purposes of
resonances and comparing the number of zeros, we focus on the dominant
modal contributions from vertical shear $A_3$ and magnetism
$(m_1,m_2)$.  Notice the general similarity between the resonant
magnetic and elastic modes in each row of
Fig.~\ref{fig:one_layer_res}, Left. Points $(a)-(c)$ correspond to an
anticrossing in the zeroth order mode, so the wave profiles $A_3$,
$m_2$ do not exhibit zeros, i.e., they are resonant. Points $(d)-(f)$
correspond to a different anticrossing originating from first order
modes, so their profiles have a single zero.  When the dispersion
curves from these two separate anticrossings intersect each other, an
additional anticrossing does not occur because the mode numbers
differ. Figures $(c)$ and $(f)$ are the magnetic profiles at
nonresonant crossings; the nonresonant correction is very small and
not visible in this figure.

Although the asymptotic calculation in section~\ref{sec:mag_asymp}
made several simplifying assumptions, its main findings are
qualitatively verified. For exchange-dominated waves, anticrossings
only appear between resonant magnetic and elastic modes. In addition,
Fig.~\ref{fig:one_layer_res} clearly shows that the higher-order
anticrossings decrease in their gap width. This is consistent with
\eqref{eq:as_comp}, which predicts that the gap width will decrease
with increasing frequency and decreasing wave number. Finally, close
observation reveals that the quasi-magnetic and quasi-elastic
dispersion curves in Fig.~\ref{fig:one_layer_res} are slightly shifted
from their uncoupled locations.

\subsection{Two layers}
\begin{figure}
  \centering
  \includegraphics[scale=.25]{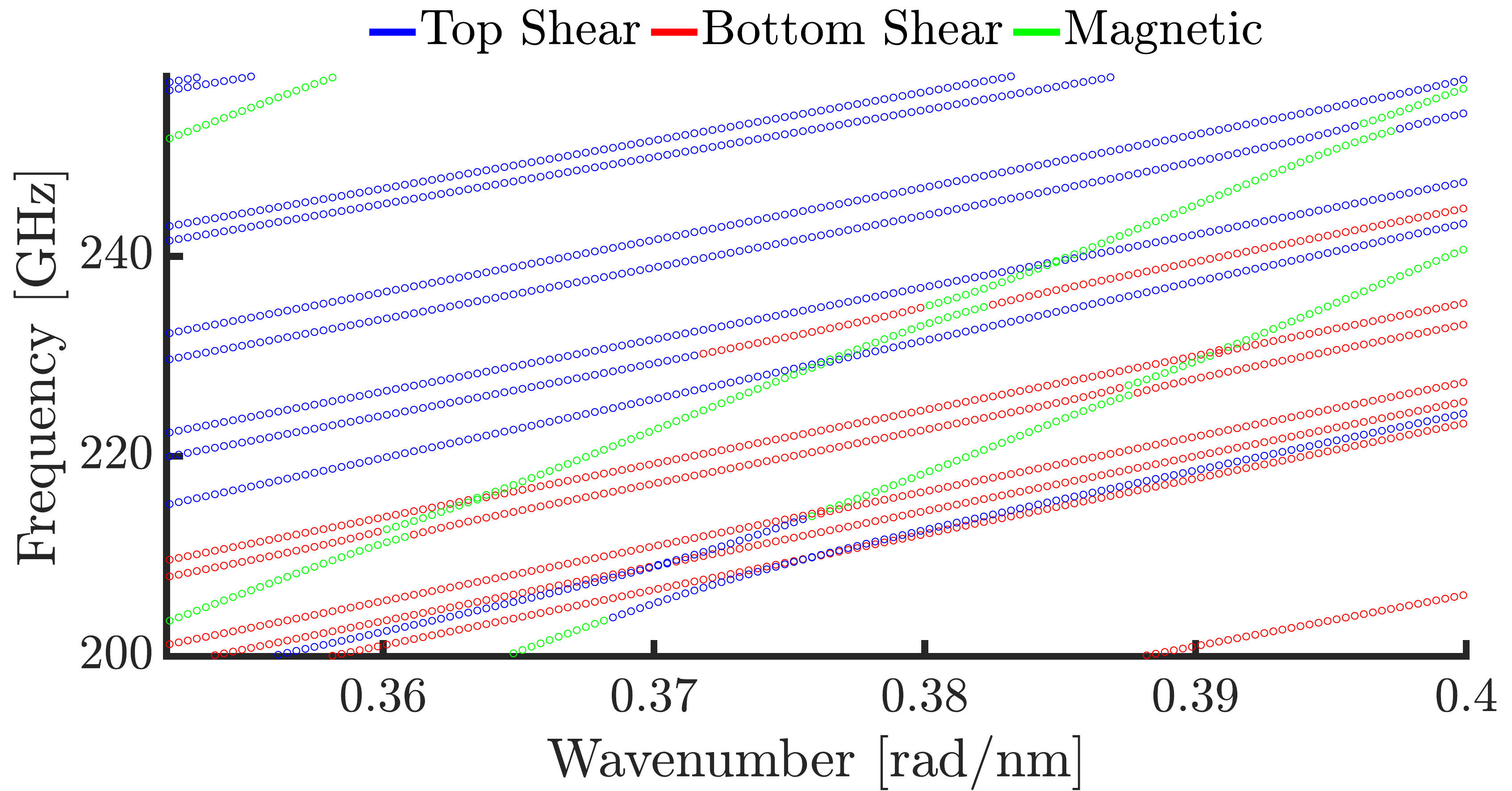}
  \caption{Dispersion curves for one layer of YIG with thickness 30 nm
    and $N=12$ layered on a 50 nm film of GGG. Numerous anticrossings
    are visible.}
  \label{fig:yig_ggg}
\end{figure}

\subsubsection{Dispersion map}
Next, we incorporate a second, nonmagnetic layer and examine its
effect on the magnetoelastic dispersion
curves. Figure~\ref{fig:yig_ggg} shows the dispersion map for 30 nm of
YIG layered on a 50 nm substrate of GGG. The plot axes were chosen to
show a large number of anticrossings. Elastic shear waves localized in
the YIG layer are colored blue, while shear waves localized in the GGG
layer are red. Magnetic waves are green. Note that there are
significantly more dispersion curves here than in
Fig.~\ref{fig:single_layer_compare}, indicating that adding a second
layer significantly increases the dispersion complexity.

Interestingly, the magnetic dispersion curves interact with elastic
modes from each layer, even though the bottom layer is nonmagnetic. In
other words, an elastic wave with energy localized in the nonmagnetic
substrate can still interact resonantly with a magnetic-dominated
mode. One consequence of this observation is that a single dispersion
curve can transition between a quasi-elastic wave in the top layer to
a quasi-elastic wave in the bottom layer or a quasi-magnetic wave,
depending on the wavenumber and frequency regime.

\begin{figure}
  \centering
  \includegraphics[scale=.25]{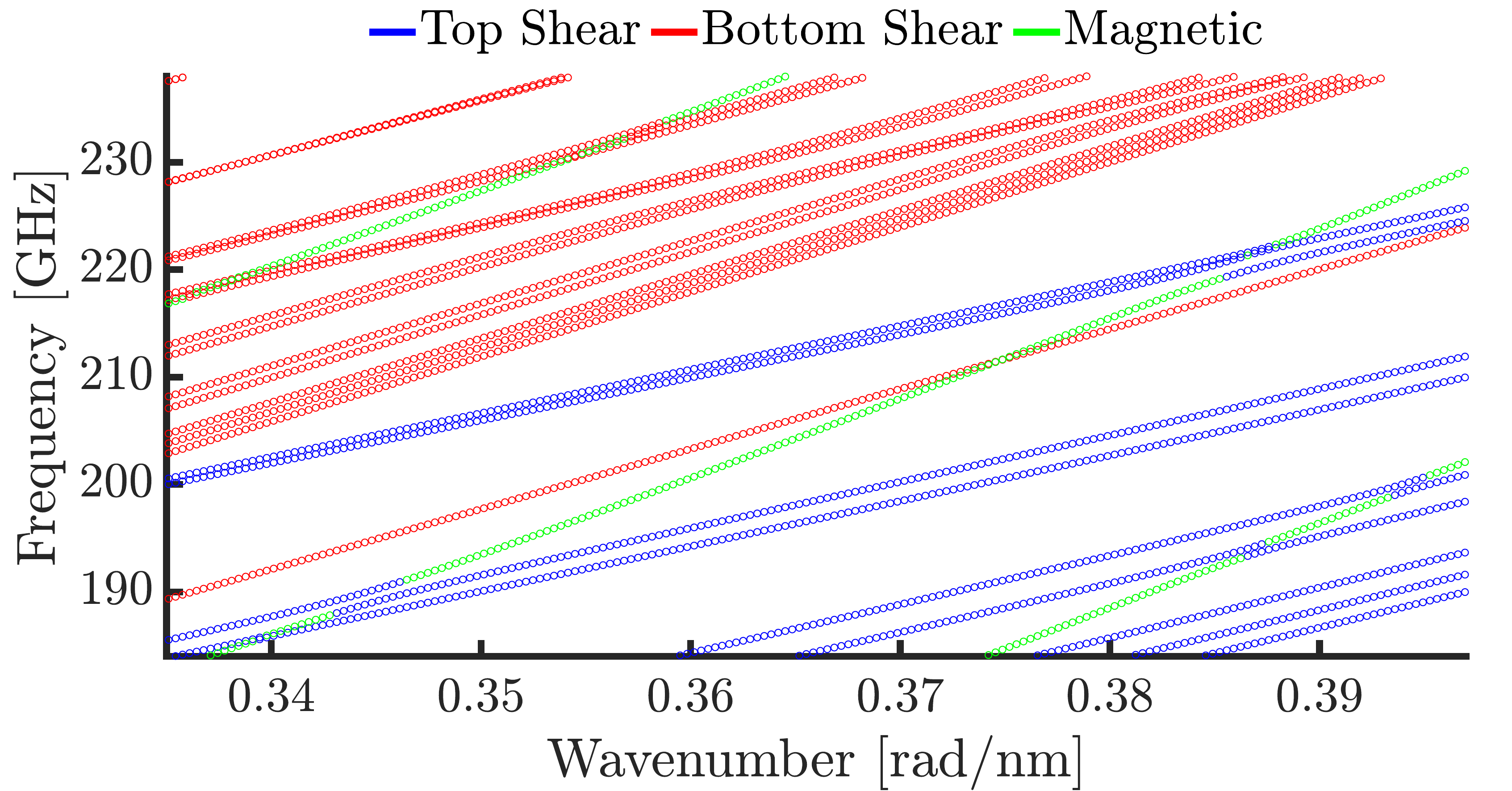}
  \caption{Dispersion curves for one one layer of Ni with thickness 50
    nm and $N=12$ layered on a 100 nm film of Si$_2$N$_3$ (right).}
  \label{fig:ni_sin}
\end{figure}

In Fig.~\ref{fig:ni_sin}, we show a dispersion map for a different
sample consisting of 50 nm of Ni layered on a 100 nm Si$_2$N$_3$
substrate. One essential difference between this calculation and that
of Fig.~\ref{fig:yig_ggg} is that here, the shear speed $c_S$ of the
bottom material is larger than that of the top material. This is
visible in the large number of dispersion modes with energy localized
in the bottom material (red curves) located at higher frequencies than
the modes with energy localized in the top material (blue curves).

Similar effects are seen here as in Fig.~\ref{fig:yig_ggg}. Modes with
energy primarily in the nonmagnetic substrate can still interact with
magnetic modes, resulting in a single dispersion branch that
transitions between all three wave types. We examine this phenomenon
in more detail next.

\subsubsection{Multimode dispersion branch in a double layer}
\begin{figure}
  \centering
  \includegraphics[scale=.25]{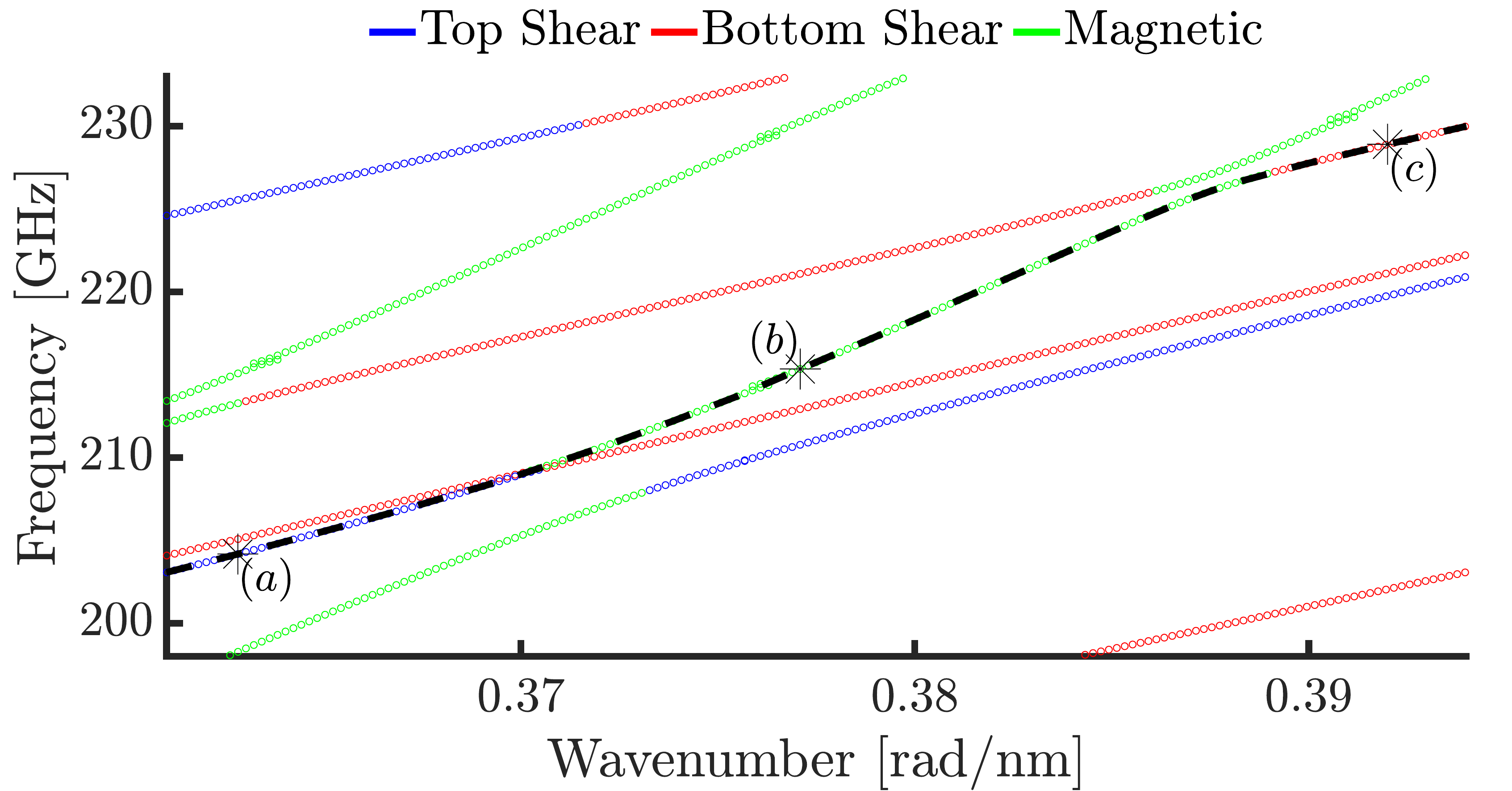}
  \includegraphics[scale=.25]{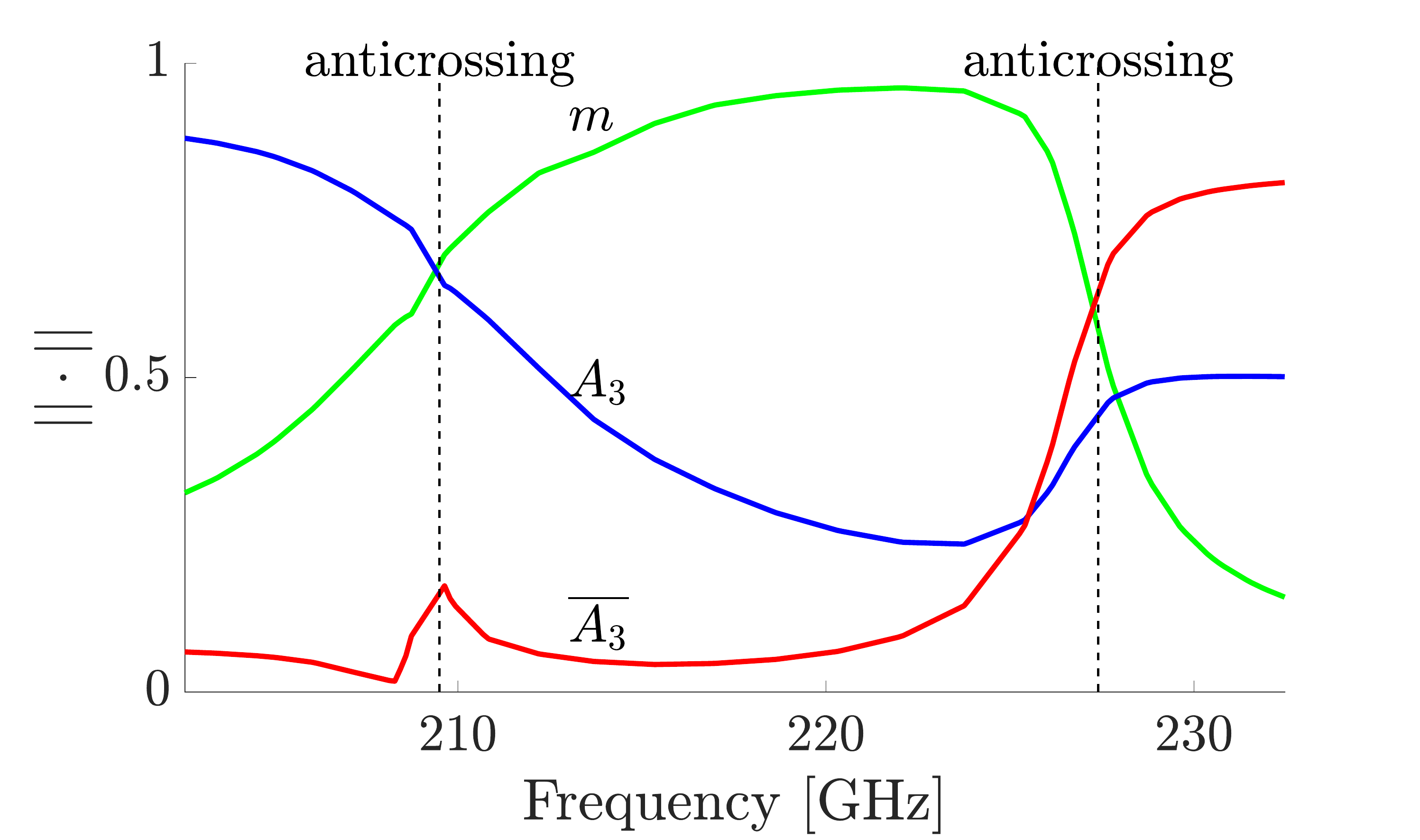}
  \caption{Top: dispersion curves for 30 nm film of YIG layered on 50
    nm of GGG with $N=24$. Bottom: relative magnitudes of the $A_3$,
    $\overline{A_3}$, and $m$ modes for the dispersion curve indicated
    by a dashed line in the top figure.}
  \label{fig:yig_ggg_zoom}
\end{figure}
Now, we continuously follow a single dispersion curve in a YIG-GGG
layered material to understand how energy can transition from magnetic
to elastic or vice-versa as well as between layers. The top panel of
Fig.~\ref{fig:yig_ggg_zoom} shows a zoomed-in view of a portion of
Fig.~\ref{fig:yig_ggg}. Dispersion curves with energy primarily
located in the $A_2$ and $\overline{A_2}$ modes have been removed for
ease of visualization since they play no role in the resonance
discussed here.

One dispersion curve in Fig.~\ref{fig:yig_ggg_zoom} has been
highlighted using a black dashed line. This curve begins as a
quasi-elastic wave localized in the top elastic layer. After an
anticrossing it becomes quasi-magnetic, and then after a second
anticrossing it becomes quasi-elastic, localized in the bottom
layer. This transition is shown more clearly in the bottom panel of
Fig.~\ref{fig:yig_ggg_zoom}, which depicts the relative magnitudes of
the three primary wave elements as the frequency is increased.

\begin{figure*}
  \centering
  \includegraphics[scale=.25]{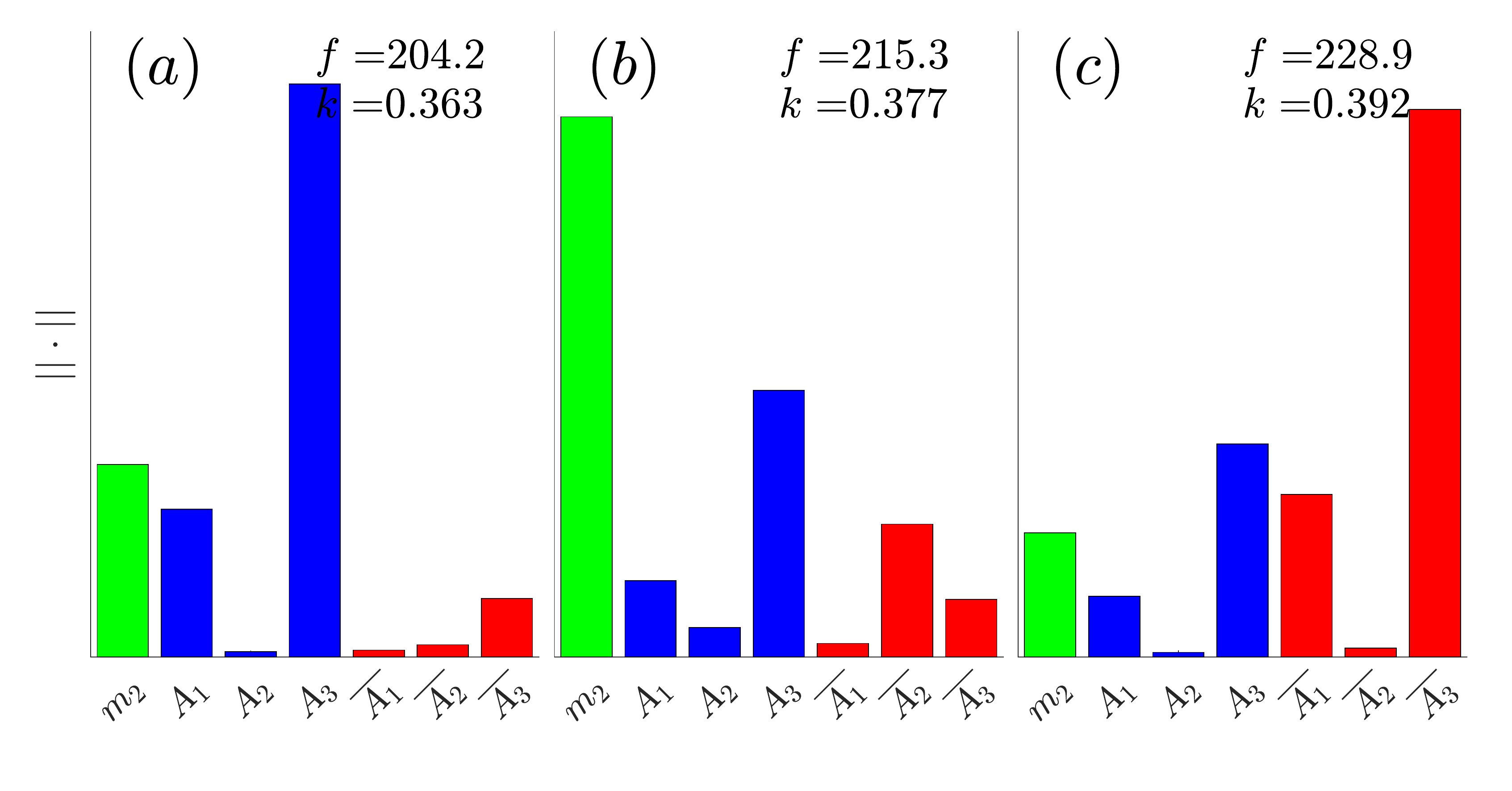} \\
  \includegraphics[scale=.25]{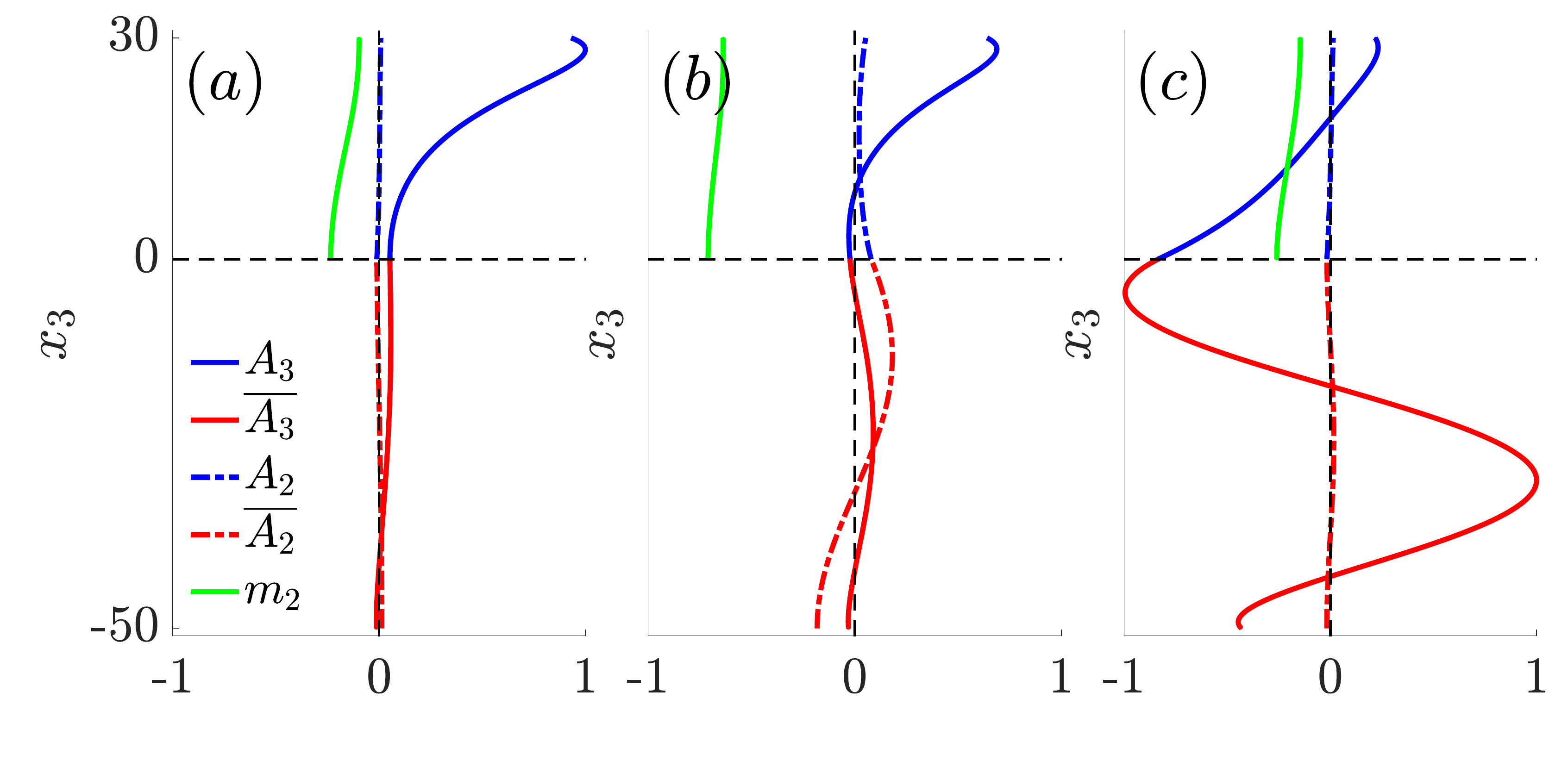} 
  \includegraphics[scale=.25]{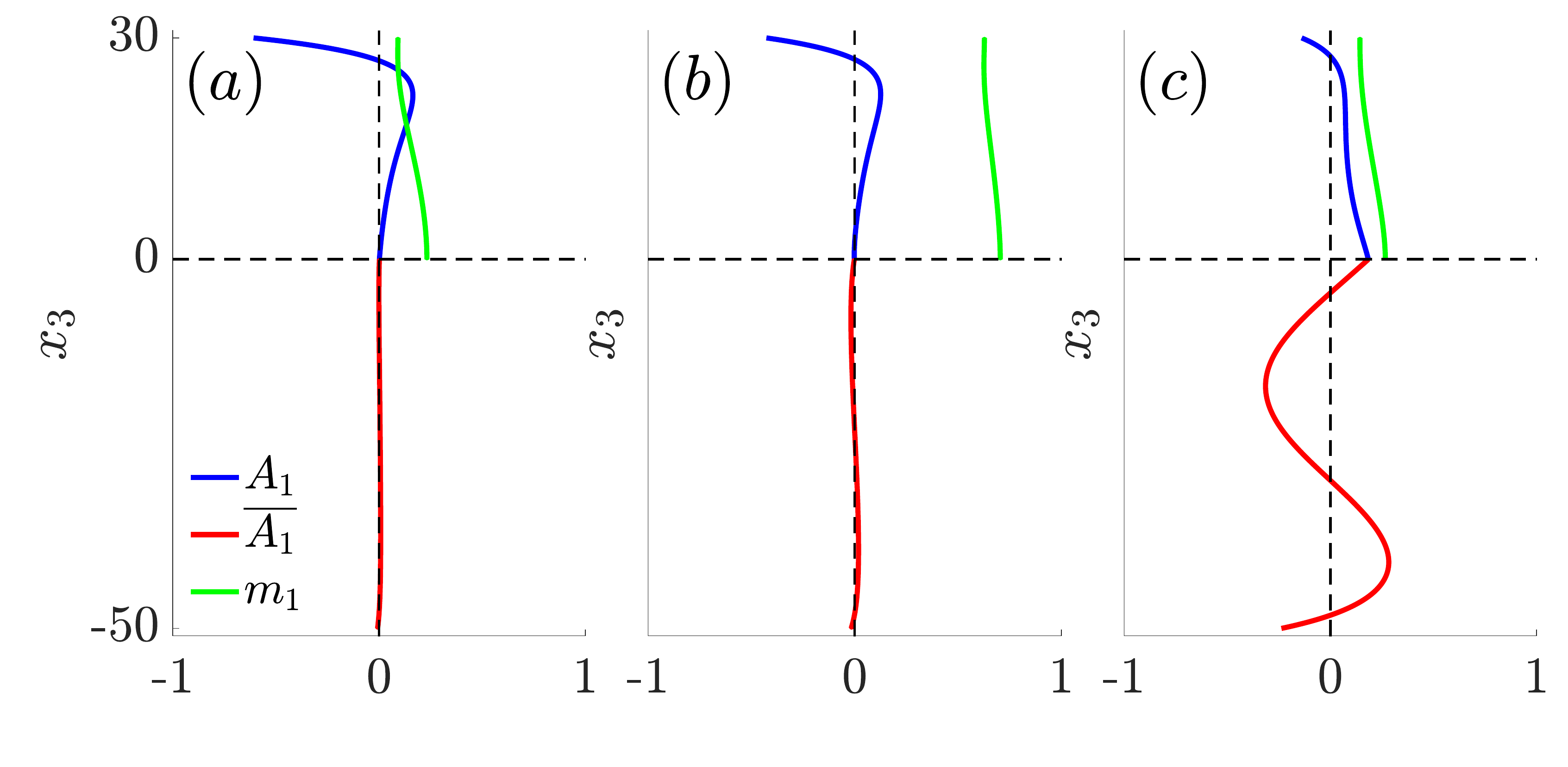}
  \caption{Top: Relative 2-norms of the scaled displacements and
    magnetic modes for the points labeled in
    Fig.~\ref{fig:yig_ggg_zoom}. Bottom: corresponding wave profiles
    of the $(A_3,\overline{A_3},A_2,\overline{A_2},m_2)$ components
    (Left) and the $\pi/2$ phase advanced $(A_1,\overline{A_1},m_1)$
    components (Right), calculated with $N=32$. The dominant
    contributions are: $(a)$ a vertically sheared surface acoustic
    wave $A_3$, $(b)$ a spin wave $(m_1,m_2)$, and $(c)$ an
    oscillatory bulk substrate mode $\overline{A_3}$.}
    \label{fig:wave_struc}
\end{figure*}

We further examine the three points labeled $(a)$, $(b)$, and $(c)$ in
Fig.~\ref{fig:wave_struc}. The top panel in Fig.~\ref{fig:wave_struc}
shows a bar graph with the relative 2-norms of all seven mode
components, suitably scaled according to \eqref{eq:nondim}, for the
three points. From this figure, it is clear that the energy in point
$(a)$ is primarily localized in the $A_3$ elastic component of the
magnetic layer, in point $(b)$ it is in the magnetic components
$(m_1,m_2)$, and in point $(c)$ it is in $\overline{A_3}$, the elastic
component of the nonmagnetic substrate.

Finally, the vertical mode profiles at the three labeled points are
shown in the bottom panel of \ref{fig:wave_struc}. At points $(a)$ and
$(b)$, the elastic portion of the wave is localized primarily at the
top surface of the magnetic layer, adjacent to vacuum.  However, when
the mode becomes dominated by vertical shear elastic oscillations in
the nonmagnetic layer, the mode is excited across much of the
nonmagnetic layer. This is accompanied by a vertical shear component
with about half the energy in the magnetic layer and a small magnetic
excitation.

Interestingly, the mode profiles for anticrossings in a layered
material do not display as simple of a resonance behavior as for a
single layer (see Fig.~\ref{fig:one_layer_res}). For example, the
magnetic mode profile in Fig.~\ref{fig:wave_struc} is order 0, the top
material elastic mode is order 0 or 1, and the bottom material elastic
mode is order 1 or 2. This causes an asymmetry with respect to the
layer center in the magnetic waves.  In contrast, this asymmetry is
not observed in single layers nor in the low frequency calculations of
the next section.  This indicates that, for a layered material,
resonant and nonresonant interactions have added complexity which is
not captured by our simplified asymptotic calculation. Nevertheless,
the presence of some anticrossings and some simple crossings is
consistent with our analytical findings.

To summarize, the highlighted dispersion curve in
Fig.~\ref{fig:yig_ggg_zoom} and corresponding figures in this section
reveal that for increasing frequency, magnetism acts as a mediator to
transfer energy between a surface shear wave in the magnetic material
to a shear wave across the nonmagnetic substrate.

\subsection{Low frequency interactions}

\begin{figure*}
  \centering
  \includegraphics[scale=.25]{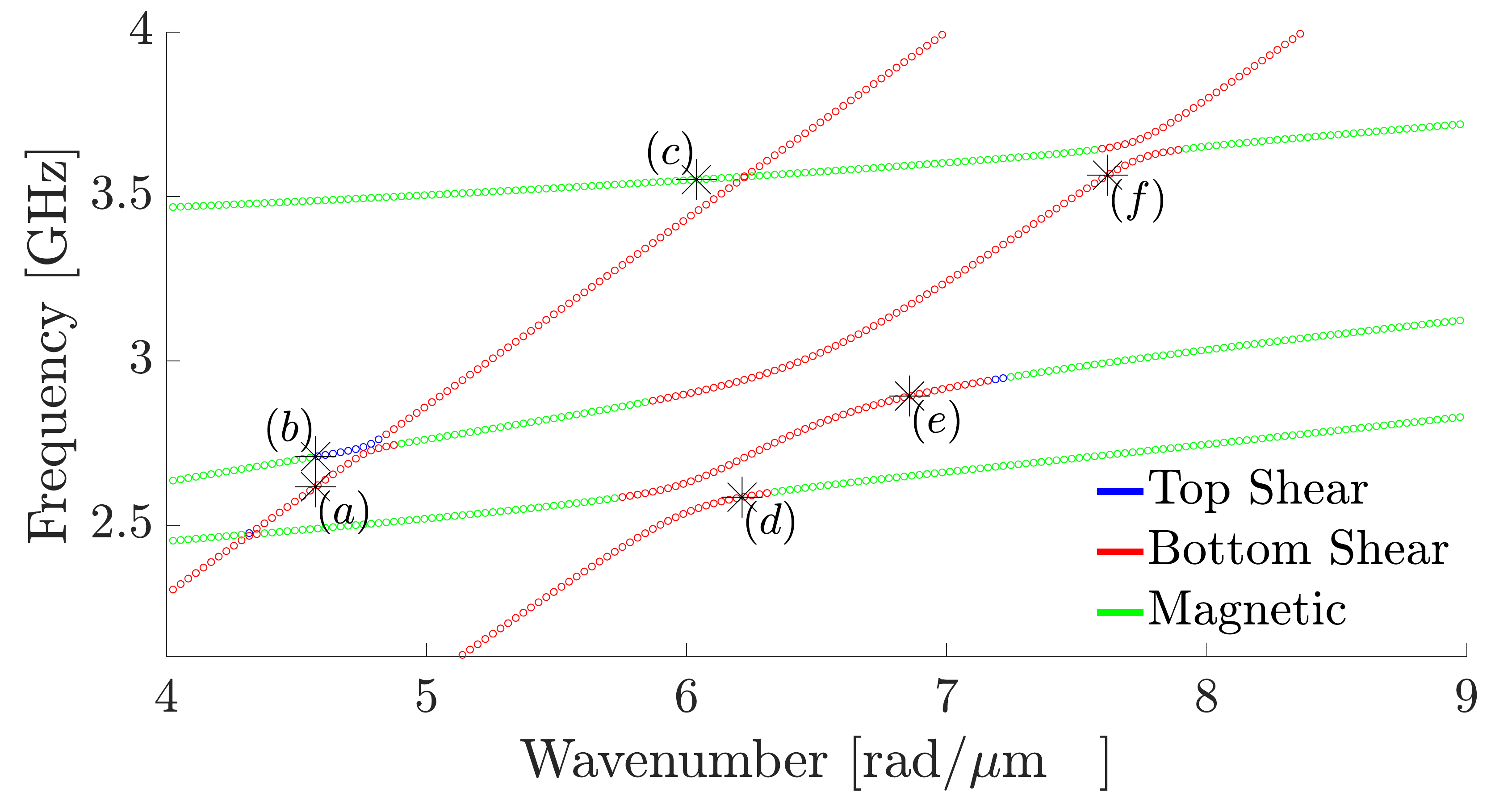}\\
  \includegraphics[scale=.25]{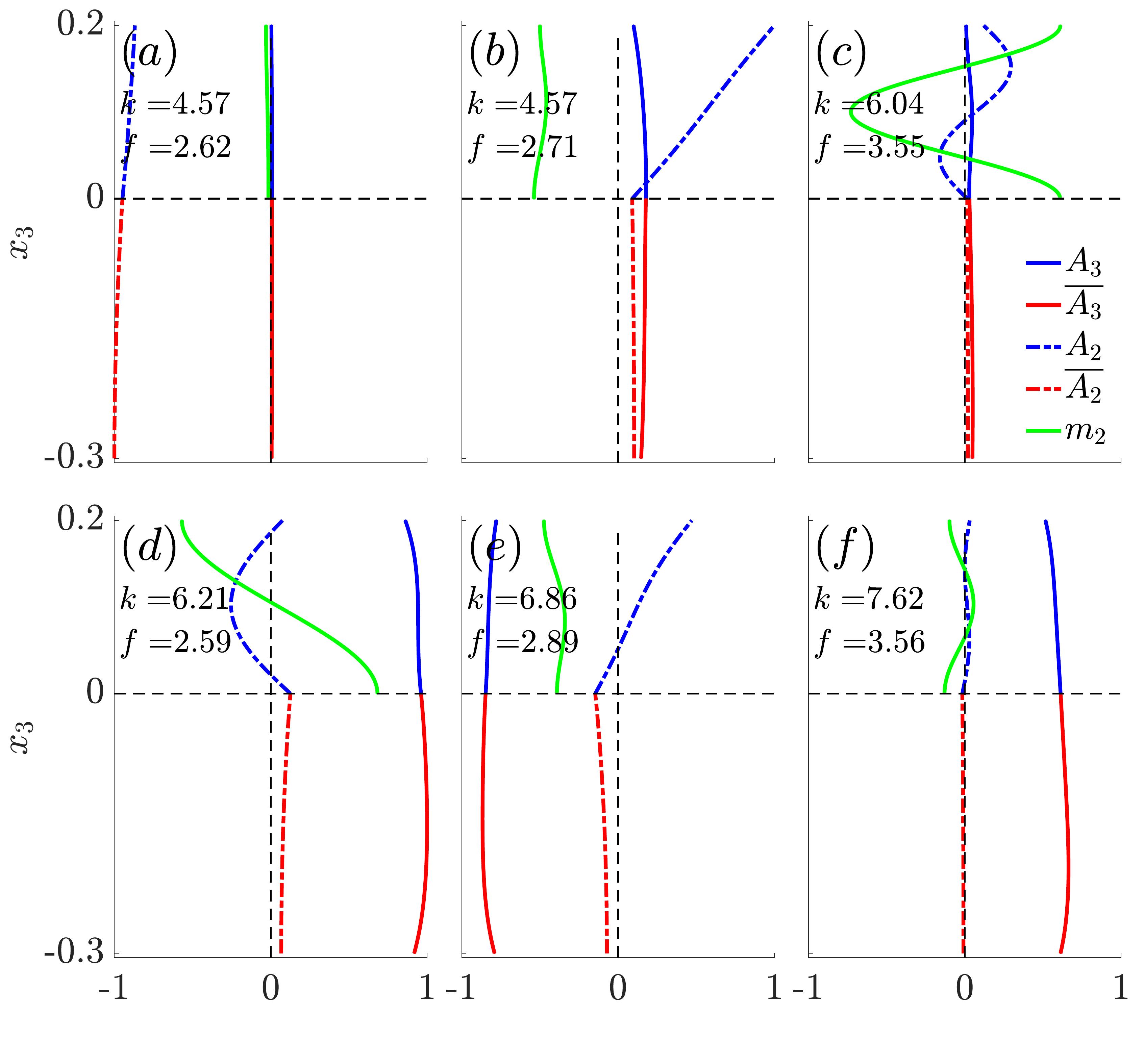} 
  \includegraphics[scale=.25]{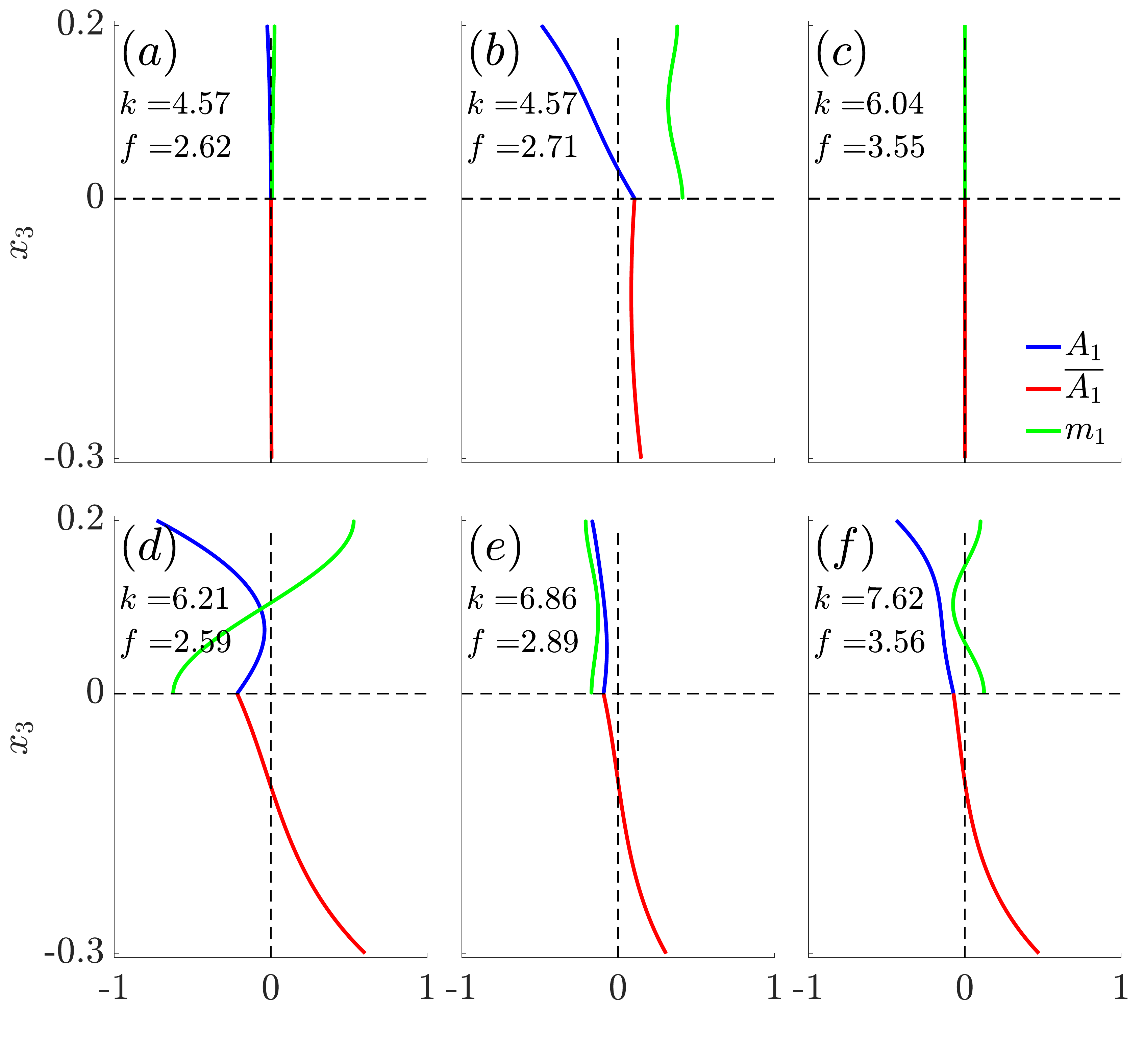}
  \caption{Top: dispersion map for 0.2 $\mu$m of YIG layered on 0.3
    $\mu$m of GGG with $N=16$. Multiple anticrossings and nonresonant
    crossings are visible. Bottom: wave profiles for the labeled
    points in the top figure. Resonant and nonresonant interactions
    are more complicated. For this figure, all wavenumbers are in
    rad/$\mu$m.}
  \label{fig:low_freq_anticrossing}
\end{figure*}

For completeness, we also present some dispersion and wave profile
results for waves in a lower-frequency range, shown in
Fig.~\ref{fig:low_freq_anticrossing}. The main phenomenological
difference for this 1--4 GHz regime is that dipole effects are much
stronger relative to exchange effects. Thus, our exchange-dominated
asymptotic calculation is not presumed to be valid here. To ensure the
presence of multiple dispersion modes, we increase the material
thicknesses to 0.2 $\mu$m of YIG layered on 0.3 $\mu$m of GGG.

Figure \ref{fig:low_freq_anticrossing} shows two quasi-elastic modes
and three quasi-magnetic modes in the dispersion map with four
resonant anticrossings and two nonresonant crossings visible. One
interesting observation from the wave profiles is that the order zero
(panels $(b)$ and $(e)$) magnetic mode appears to track the middle
quasi-magnetic dispersion curve, with higher frequency than the first
order magnetic mode in panel $(d)$.

The leftmost quasi-elastic dispersion curve (panel $(a)$) appears to
be a transverse shear, zeroth-order elastic mode with energy
concentrated in $A_2$, while the rightmost quasi-elastic curve (panels
$(d)$, $(e)$, $(f)$) is also a zeroth-order mode but with energy
concentrated in vertical shear $A_3$. The $A_2$ mode only has an
anticrossing with the second magnetic mode, which from panels $(b)$
and $(e)$ also appears to be a zeroth-order mode. This is generally
consistent with the asymptotic results obtained earlier regarding
resonant and nonresonant anticrossings.

However, the $A_3$ mode (panels $(d)$, $(e)$, and $(f)$) displays
anticrossing behavior with all three magnetic modes. Apparently, the
presence of strong dipole effects alters whether magnetoelastic
intersections are resonant or nonresonant. Since the dipole effects
occur in the vertical $x_3$-direction, this may explain why resonant
and nonresonant interactions are still present for the $A_2$-mode,
which does not experience as strongly the presence of the dipole
effective field.

\section{Discussion and conclusion}
\label{sec:disc-concl}

The main contribution of this work is an analysis of magnetoelastic
dispersion in finitely thick and layered materials. We first presented
fully coupled magnetoelastic equations, which were then linearized to
obtain a coupled system of ordinary differential equations in the
vertical, layered direction. This system was analyzed analytically and
numerically.

Analytically, we performed an asymptotic calculation to study multiple
dispersion curve crossings and anticrossings in a single layer. We
identified that anticrossings only occur when the uncoupled magnetic
and elastic dispersion curves are resonant with each other. These
anticrossings were well-modeled using asymptotic predictions. We also
correctly predicted a small but noticeable correction for nonresonant
dispersion curve intersections from the uncoupled case. Despite
performing our calculations on a simplified scenario, these same
general behaviors were observed numerically for the full equations in
a single layer. The calculation assumed weak magnetoelastic coupling,
which is the case for all known materials.

Numerically, we introduced a magnetoelastic spectral collocation
method (SCM), a fast, simple, accurate, direct approach for
calculating the dispersion curves of a magnetoelastic, layered
material. Discretizing the coupled system of differential equations at
Chebyshev points yields a polynomial eigenvalue problem whose
solutions are the angular frequencies for a given wavenumber. Solving
this polynomial eigenvalue problem for a mesh of wavenumbers gives
both dispersion curves and corresponding wave profiles.

The speed and reliability of SCM was first validated in multiple
ways. We then applied SCM to a variety of materials and
samples. First, we calculated the dispersion curves for a 30 nm layer
of YIG surrounded by a vacuum. This dispersion map revealed a number
of anticrossings and simple crossings. Examining the corresponding
wave structures supported our analytical result that, for a single
layer, anticrossings only occur for resonant magnetic and elastic
waves.

Next, we calculated dispersion curves for two different double
layers. Both calculations showed complex dispersion maps with numerous
anticrossings. In addition, it was observed that the same dispersion
curve can transition between quasi-elastic in the top layer,
quasi-magnetic, and then quasi-elastic in the bottom, nonmagnetic
layer. The magnetic curve is resonant with an elastic curve in both
the bottom and top layers, and so mediates energy transfer between
them with increasing frequency. This curve was subsequently
investigated by examining the wave structures at various points.

We also found that resonant and nonresonant interactions are
signficantly complicated by dipole effects and layering. Both effects
can lead to results that are at odds with our simplified analytical
predictions for a single layer and negligible dipole field.  This
highlights the importance of the numerical method for determining the
solutions to the full linearized system of differential equations.

We emphasize the fact that SCM is easily generalizable to many
magnetoelastic wave applications. The method works just as effectively
for GHz frequency ranges as for the extremely high frequency ranges
primarily considered here. Calculations for a material with an
in-plane magnetic field are also readily obtainable. In that case, the
linearization assumptions would lead to a slightly different set of
differential equations, but these can be similarly discretized and
converted into a polynomial eigenvalue problem. The implementation is
straightforward, and spectral convergence ensures fast computational
times. In addition, SCM is also applicable to more complex
anisotropies and geometries, such as a spherical material or a
material with more layers. For example, experimental work reported in
\cite{an_momentum_2020} studied magnetoelastic waves in a layer of GGG
between two layers of YIG. Although the computational complexity would
be increased by the need to incorporate dipole-dipole interactions,
SCM could be adapted to this scenario to provide valuable insights.

Another advantage of SCM is the ability to recover wave profiles with
spectral accuracy. These profiles can then be utilized to classify the
dispersion mode type. In addition, we can identify transitions along a
dispersion curve between the various wave types. This ability to
recover wave profiles could also lead in some interesting
directions. Ultrafast magnetism experiments measuring scattering
intensities using an XFEL yield only a wavenumber-frequency relation;
they contain no information about the wave type or structure. By
fitting experimentally-obtained dispersion curves to SCM results, one
can predict information about the wave structure. Similarly, wave
structure data can also aid in the calculation of surface waves such
as Rayleigh or Love modes, or waves localized to an interface, another
area with many practical applications.

\begin{acknowledgments}
  The work of SR and MAH was partially supported by the
  U.S.~Department of Energy, Office of Science, Grant
  No.~DE-SC0018237.  The authors are grateful to Thomas J.~Silva for
  suggesting this problem and many inspiring discussions.  SR also
  thanks Danny Ramasawmy for helpful communication in the development
  of this method.
\end{acknowledgments}

\medskip

\appendix

\section{Asymptotic calculation of exchange-dominated
  magnetoelastic dispersion}
\label{sec:asympt-calc-exch}

\renewcommand{\thefigure}{A\arabic{figure}}
\setcounter{figure}{0}

Setting $A_1$, $A_2$, and $g$ to zero in the eigenvalue problem
\eqref{eq:full_coupled_system} leaves three equations for $m_1$,
$m_2$, and $A_3$.  These can be further simplified by solving
\eqref{eq:mag_1} for $m_1$ in terms of $m_2$, differentiating twice,
and inserting into \eqref{eq:mag_2} so that $m_1$ is eliminated. If we
write $m \equiv m_2$ and $A \equiv A_3$ for convenience, then the new,
simplified eigenvalue problem is sixth order
\begin{subequations}
  \begin{widetext}
    \label{eq:54}
    \begin{align}
      (\omega^2-c_S^2k^2)A+c_L^2 A'' - \frac{k B_2}{\omega \rho M_{\rm
      s}}[(\omega_{\rm H}-\omega_{\rm M} +\beta k^2)m-\beta m''] &= 0,
      \\   
      [\omega^2-(\omega_{\rm H}-\omega_{\rm M}+\beta k^2)^2
      ]m+2\beta (\omega_{\rm H}-\omega_{\rm M}+\beta k^2) m'' -
      \beta^2 m'''' - \omega k \gamma B_2 A &= 0, 
    \end{align}
  \end{widetext}
  for $0<x_3<d$. Recall the definitions of $\omega_{\rm H}$,
  $\omega_{\rm M}$ and $\beta$ in \eqref{eq:2}.  The corresponding six
  boundary conditions are
  \begin{equation}
    A'\equiv m' \equiv m''' \equiv 0, \qquad x_3 = 0, \; x_3 = d.
  \end{equation}
\end{subequations}

It will be helpful to nondimensionalize \eqref{eq:54}.  For this, we
introduce the scalings
\begin{subequations}
  \label{eq:nondim}
  \begin{equation}
    \label{eq:5}
    \begin{split}
      \tilde{k} &= k/K, \quad \tilde{\omega} = 
                  \omega/\Omega, \quad \tilde{x}_3 = K x_3, \\
      \tilde{m} &= m/M_{\rm s}, \quad \tilde{A} = A/A_* ,
    \end{split}
  \end{equation}
  where all nondimensional quantities are distinguished from their
  dimensional counterpart by a tilde $\tilde{\ }$.  Furthermore, the
  wavenumber, frequency, and elastic displacement scalings are
  \begin{equation}
    \label{eq:60}
    K = c_S/\beta, \quad \Omega = c_S K, \quad A_* = \sqrt{\frac{M_{\rm
          s}\beta}{\rho c_S^2 \gamma}}.
  \end{equation}
  The lowest order magnetoelastic intersection occurs for the
  spatially uniform elastic mode; hence the frequency scaling
  $\Omega = c_S K$.  The wavenumber $K$ is determined by equating
  elastic and quadratic exchange dispersion $c_S K = \beta K^2$,
  determining a natural length scale for the waves in this problem.
  The magnetization $M_{\rm s}$ and displacement $A_*$ scalings
  provide the means to directly compare magnetic and elastic mode
  amplitudes.  This scaling is used in all mode comparison plots:
  Figs.~\ref{fig:one_layer_res}, \ref{fig:yig_ggg_zoom},
  \ref{fig:wave_struc}, and \ref{fig:low_freq_anticrossing}.  The
  scalings in \eqref{eq:nondim} imply that $\tilde{\omega}=\tilde{k}=1$
  when the lowest order uncoupled dispersion branches intersect
  $\omega_{\rm H}=\omega_{\rm M}$.  Under the above scalings, we
  obtain the nondimensional applied field parameter $\tilde{H}$ and
  elasticity constant $\tilde{G}$,
  \begin{equation}
    \label{eq:6}
    \tilde{H} = \frac{(\omega_{\rm H}-\omega_{\rm M})\beta}{c_S^2},  \qquad
    \tilde{G} = \frac{c_L}{c_S}. 
  \end{equation}
\end{subequations}
Finally, we introduce the nondimensional magnetoelastic coupling
parameter $\epsilon$ in eq.~\eqref{eq:epsilon}.  Then, upon using
\eqref{eq:epsilon} and \eqref{eq:nondim}, the eigenvalue problem
\eqref{eq:54} becomes
\begin{subequations}
  \label{eq:64}
  \begin{align}
    \label{eq:64a}
    (\tilde{\omega}^2-\tilde{k}^2)\tilde{A}+\tilde{G}^2 \tilde{A}''- \epsilon
    \frac{\tilde{k}}{\tilde{\omega}} 
    [(\tilde{H}+\tilde{k}^2) \tilde{m}-\tilde{m}''] &= 0, \\ 
    \label{eq:64b}
    [\tilde{\omega}^2-(\tilde{H}+\tilde{k}^2)^2]\tilde{m}+2(\tilde{H}+
    \tilde{k}^2) \tilde{m}'' -  \tilde{m}'''' - 
    \epsilon \tilde{\omega} \tilde{k} \tilde{A} &= 0, 
  \end{align}
  for $0<\tilde{x}_3<\tilde{d}=\frac{c_{S} d}{\beta}$. For nickel,
  $\epsilon \approx 0.019$, while for YIG $\epsilon \approx 0.023$,
  both small parameters (see Table~\ref{table:speeds}). The two
  physical parameters $\tilde{H}$ and $\tilde{G}$ are
  $\mathcal{O}(1)$, as are $\tilde{\omega}$ and $\tilde{k}$. Under the
  above transformation, the boundary conditions become
  \begin{equation}
    \label{eq:64c}
    \tilde{A}'\equiv \tilde{m}' \equiv \tilde{m}''' \equiv 0, \qquad
    \tilde{x}_3 = 0, \; \tilde{x}_3 = \tilde{d}.
  \end{equation}
\end{subequations}
For a layer of nickel with thickness $d=50$ nm,
$\tilde{d}\approx 24 \gg 1$.
We now perform an asymptotic analysis of \eqref{eq:64} as
$\epsilon \to 0^+$.

\subsection{Zeroth order solution}
First, we consider the uncoupled equations. Setting $\epsilon = 0$ in \eqref{eq:64} yields
the mode profiles
\begin{subequations}
  \label{eq:fund_soln}
  \begin{align}
    \label{eq:fund_soln_a}
    \tilde{A}_n(\tilde{x}_3) &= \tilde{a}_{{\rm el},n}\cos
                               (\tilde{\xi}_n \tilde{x}_3 ), \quad n =
                               0, 1, \ldots, \\
    \label{eq:fund_soln_m}
    \tilde{m}_j(\tilde{x}_3) &= \tilde{a}_{{\rm m},j} \cos
                               (\tilde{\xi}_j \tilde{x}_3), \quad j =
                               0, 1, \ldots ,
  \end{align}
\end{subequations}
where $\tilde{\xi}_n = n\pi/\tilde{d}$ is the vertical wavenumber.
The uncoupled dispersion relations for these modes are
\begin{subequations}
  \label{eq:uncoupled_dispersion}
  \begin{align}
    \label{eq:el_uncoup_disp}
    \tilde{\omega}_n(\tilde{k})&=\sqrt{\tilde{k}^2+\tilde{G}^2\xi_n^2},\quad n =
                                 0,1,2,\dots, \\ 
    \label{eq:mag_uncoup_disp}
    \tilde{\omega}_j(\tilde{k}) &=\tilde{H}+ \tilde{k}^2 + \xi_j^2, \quad \, j =
                                  0,1,2,\dots .
  \end{align}
\end{subequations}
Note that there are actually two dispersion branches for each
component $\pm \tilde{\omega}$, but we will focus on the two positive
branches in \eqref{eq:el_uncoup_disp}, \eqref{eq:mag_uncoup_disp}.
These are the dispersion curves for a single layer. As
$\tilde{d} \to \infty$, the dispersion curves converge to the bulk
dispersion curves in which $\tilde{\xi}_n,\tilde{\xi}_j \to 0$.

\subsection{Quasi-elastic and quasi-magnetic waves}

Next, we consider the effect of coupling on quasi-elastic and
quasi-magnetic dispersion. We expand around the uncoupled modes
\eqref{eq:fund_soln} and dispersion branches
\eqref{eq:uncoupled_dispersion} using the weak coupling parameter
$\epsilon$. First, we expand around a purely elastic wave, assuming a
weak magnetic component
\begin{equation}
  \label{eq:1}
  \begin{split}
    \tilde{\omega} &\sim \tilde{\omega}_n^{(0)} + \epsilon
                     \tilde{\omega}_n^{(1)} + \epsilon^2
                     \tilde{\omega}_n^{(2)},  \\
    \tilde{A}_n(\tilde{x}_3) &\sim \tilde{A}_n^{(0)}(\tilde{x}_3) +
                               \epsilon
                               \tilde{A}_n^{(1)}(\tilde{x}_3)+\epsilon^2
                               \tilde{A}_n^{(2)}(\tilde{x}_3), \\
    x
    \tilde{m}_n(\tilde{x}_3) &\sim \epsilon
                               \tilde{m}_n^{(1)}(\tilde{x}_3)+\epsilon^2
                               \tilde{m}_n^{(2)}(\tilde{x}_3), 
    \end{split}
\end{equation}
for $0 < \epsilon \ll 1$. The zeroth order mode $\tilde{A}_n^{(0)}$
corresponds to the profile in \eqref{eq:fund_soln_a}, while
$\tilde{\omega}_n^{(0)}$ corresponds to the uncoupled elastic
dispersion frequency branch of order $n$
\eqref{eq:el_uncoup_disp}. Then, at $\mathcal{O}(\epsilon)$,
eq.~\eqref{eq:64} becomes
\begin{widetext}
  \begin{subequations}
    \label{eq:100}
    \begin{align}
      \label{eq:100a}
     \tilde{\omega}_n^{(0)} [(\tilde{\omega}_n^{(0)})^2 -\tilde{k}^2]
      \tilde{A}_n^{(1)} + \tilde{\omega}_n^{(0)} \tilde{G}^2
      (\tilde{A}_n^{(1)})''
      &= F_1 , \\
      \label{eq:100b}
      [(\tilde{\omega}_n^{(0)})^2-(\tilde{H}+\tilde{k}_*^2)^2]\tilde{m}_n^{(1)}+2
      (\tilde{H}+\tilde{k}^2) (\tilde{m}_n^{(1)})'' -
      (\tilde{m}_n^{(1)})'''' &= G_1,
    \end{align}
  \end{subequations}
\end{widetext}
with right hand sides
\begin{equation}
  \label{eq:7}
  \begin{split}
    F_1 &= -\tilde{\omega}_n^{(1)} \tilde{A}_n^{(0)}
    [3(\tilde{\omega}_n^{(0)})^2 - \tilde{k}^2  
          - \tilde{G}^2\tilde{\xi}_n^2 ], \\
    G_1 &= \tilde{k} \tilde{\omega}_n^{(0)} 
    \tilde{A}_n^{(0)}.
  \end{split}
\end{equation}
The orthogonality (solvability) condition for
$\tilde{A}_n^{(1)}(\tilde{x}_3)$ in \eqref{eq:100a} (see, e.g.,
\cite{hinch_perturbation_1991}) implies that
$\tilde{\omega}_n^{(1)}=0$. Thus, \eqref{eq:100a} is a homogeneous
equation, and the particular solution is $\tilde{A}_n^{(1)}\equiv
0$. We can solve \eqref{eq:100b} for $\tilde{m}_n^{(1)}(\tilde{x}_3)$
as
\begin{equation}
  \label{eq:quasiel_mag}
  \begin{split}
    \tilde{m}_n^{(1)}(\tilde{x}_3) &= \tilde{b}_n \tilde{a}_{{\rm el},n}
                                     \cos(\tilde{\xi}_n \tilde{x}_3), \\
    \tilde{b}_n &= \frac{\tilde{\omega}_n^{(0)}
                  \tilde{k}} {(\tilde{\omega}_n^{(0)})^2
                  -(\tilde{H}+\tilde{k}^2)^2 
                  -2\tilde{\xi}_n^2(\tilde{H}
                  +\tilde{k}^2)-\tilde{\xi}_n^4}.
  \end{split}
\end{equation}
Importantly, the solution \eqref{eq:quasiel_mag} only holds when the
asymptotic expansion \eqref{eq:1} remains well-ordered, i.e.,
$\tilde{b}_n$ is at most $\mathcal{O}(1)$.  Consequently, when the
denominator of $\tilde{b}_n$ is small
\begin{equation}
  \label{eq:el_resonance}
  (\tilde{\omega}_n^{(0)})^2 -(\tilde{H}+\tilde{k})^2 -2\tilde{\xi}_n^2
  (\tilde{H}+\tilde{k}^2) - \tilde{\xi}_n^4  \leq
  \mathcal{O}(\epsilon),
\end{equation}
we obtain the precise statement of the resonance condition
\eqref{eq:resonance}.  In this case, we require an alternative
asymptotic expansion. Equation \eqref{eq:el_resonance} is satisfied
when the elastic curve of order $n$ \eqref{eq:el_uncoup_disp}
intersects a magnetic dispersion curve of the same order $j = n$
\eqref{eq:mag_uncoup_disp}. Thus, the asymptotic expansion
\eqref{eq:1} is only valid far from these resonant interactions. It is
not valid when magnetic and elastic dispersion curves are in
resonance, a case which we describe below.

In order to determine the higher-order frequency correction
$\tilde{\omega}_n^{(2)}$, we continue on to $\mathcal{O}(\epsilon^2)$,
which yields the same equations as in \eqref{eq:100} but with
$\tilde{A}_n^{(1)} \to \tilde{A}_n^{(2)}$,
$\tilde{m}_n^{(1)} \to \tilde{m}_n^{(2)}$ and the alternative right
hand sides
\begin{equation}
  \label{eq:8}
  \begin{split}
    F_2 &= \tilde{k} [(\tilde{H}+\tilde{k}^2) \tilde{m}_j^{(1)} -
          (\tilde{m}_j^{(1)})''] -
          2\tilde{A}_n^{(0)} (\tilde{\omega}_n^{(0)})^2
          \tilde{\omega}_n^{(2)}, \\ 
    G_2 &= \tilde{k} \tilde{\omega}_n^{(0)} \tilde{A}_n^{(1)}.
  \end{split}
\end{equation}
Since the elastic correction
$\tilde{A}_n^{(1)}(\tilde{x}_3) \equiv 0$, we obtain
$\tilde{m}_j^{(2)}(\tilde{x}_3)=0$. After inserting the solution for
$\tilde{m}_j^{(1)}(\tilde{x}_3)$ in \eqref{eq:quasiel_mag} into
\eqref{eq:8} and requiring solvability for $\tilde{A}_n^{(2)}$, we
obtain the second order dispersion curve correction
\begin{equation}
  \label{eq:el_correction}
  \tilde{\omega}_n^{(2)} = \frac{ B_n}{2 (\tilde{\omega}_n^{(0)})^2  }
  (\tilde{H}+\tilde{k}^2+\tilde{\xi}_n^2). 
\end{equation}
Thus, a quasi-elastic dispersion curve, away from resonant
intersections with a magnetic dispersion curve, has the
$\mathcal{O}(\epsilon)$ magnetic component \eqref{eq:quasiel_mag}, and
the $\mathcal{O}(\epsilon^2)$ shift in frequency
\eqref{eq:el_correction}.

We can similarly expand around a quasi-magnetic dispersion curve as
\begin{equation}
  \label{eq:9}
  \begin{split}
    \tilde{\omega} &\sim \tilde{\omega}_j^{(0)} + \epsilon
                     \tilde{\omega}_j^{(1)} + \epsilon^2
                     \tilde{\omega}_j^{(2)}, \\
    \tilde{A}_j(\tilde{x}_3) &\sim \epsilon \tilde{A}_j^{(1)}
                               (\tilde{x}_3) + \epsilon^2
                               \tilde{A}_j^{(2)}(\tilde{x}_3), \\
    \tilde{m}_j(\tilde{x}_3) &\sim \tilde{m}_j^{(0)}(\tilde{x}_3) +
                               \epsilon \tilde{m}_j^{(1)}(\tilde{x}_3)
                               + \epsilon^2 \tilde{m}_j^{(2)}(\tilde{x}_3).
  \end{split}
\end{equation}
Here, the zeroth order solution $\tilde{m}_j^{(0)}(\tilde{x}_3)$
corresponds to \eqref{eq:fund_soln_m}, while $\tilde{\omega}_j^{(0)}$
corresponds to \eqref{eq:mag_uncoup_disp}. A similar calculation to
the quasi-elastic case yields $\tilde{\omega}_j^{(1)}=0$,
$\tilde{m}_j^{(1)}=0$, and
\begin{equation}
  \label{eq:quasimag_el}
  \begin{split}
    \tilde{A}_j^{(1)}(\tilde{x}_3) &= C_j \tilde{m}_j \cos(\tilde{\xi}_j
                                     \tilde{x}_3), \\
    C_j &= \frac{\tilde{k}}{\tilde{\omega}_j^{(0)}
          [(\tilde{\omega}_j^{(0)})^2-\tilde{k}^2-\tilde{G}^2
          \tilde{\xi}_j^2]},
  \end{split}
\end{equation}
as long as the resonance condition 
\begin{equation}
  \label{eq:mag_resonance}
  (\tilde{\omega}_j^{(0)})^2- \tilde{k}^2-\tilde{G}^2 \tilde{\xi}_j^2
  \leq \mathcal{O}(\epsilon)
\end{equation}
is not met. Note that \eqref{eq:mag_resonance} is identical to
\eqref{eq:el_resonance} when $n=j$. A similar solvability condition at
$\mathcal{O}(\epsilon^2)$ yields the higher-order correction in the
quasi-magnetic frequency,
\begin{equation}
  \label{eq:mag_correction}
  \tilde{\omega}_j^{(2)} = \frac{\tilde{k}C_j}{2}.
\end{equation}

\begin{figure}
  \centering
  \includegraphics[scale=.25]{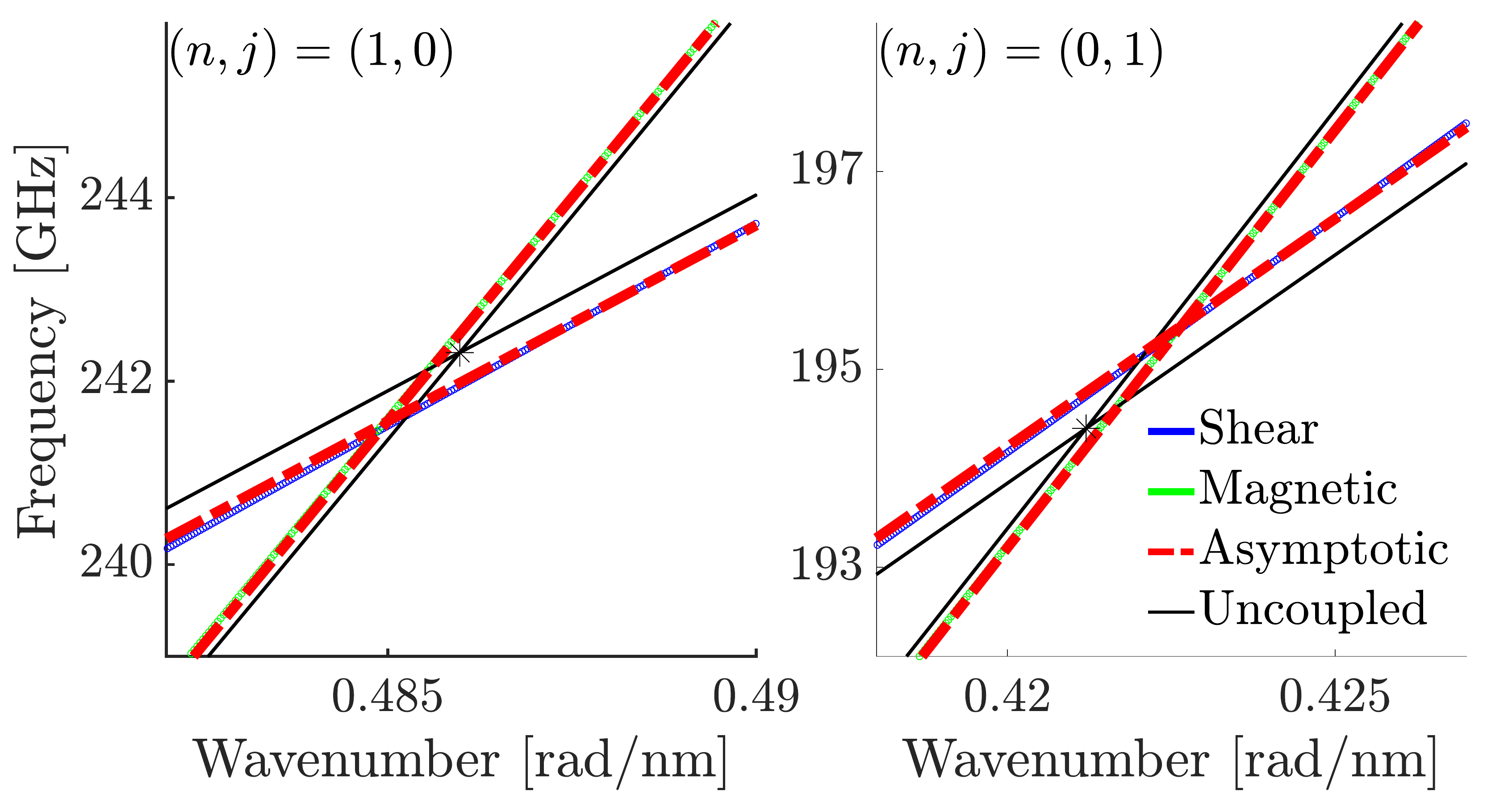} \\[1mm]
  \includegraphics[scale=.25]{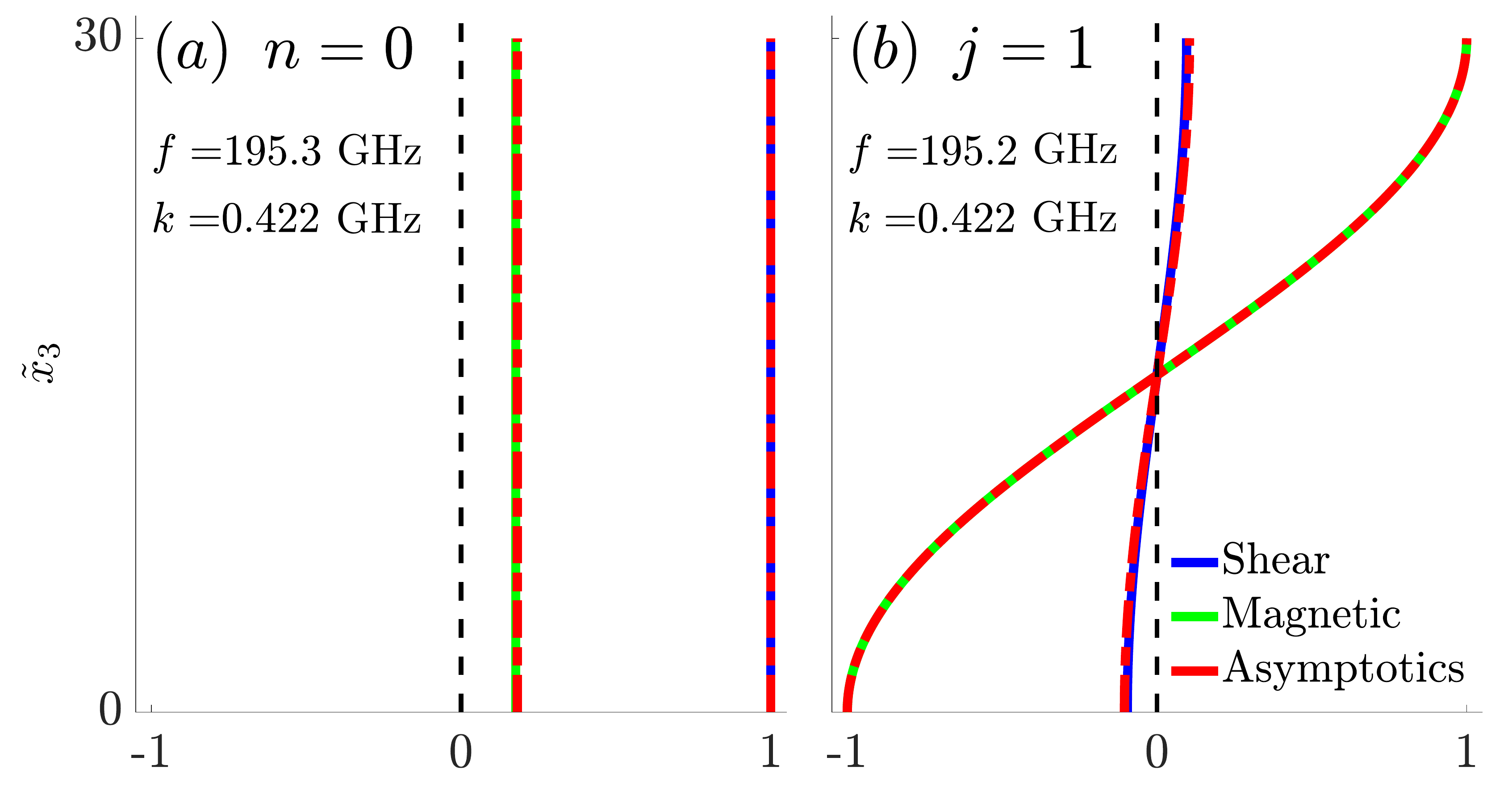}
  \caption{Top: two figures comparing the dimensional asymptotic
    prediction (dash-dotted) with results from the SCM method
    (solid). The uncoupled dispersion (solid) is also plotted to show
    that the coupled dispersion curves differ slightly. Bottom:
    comparison of the asymptotic predictions (dash-dotted)
    \eqref{eq:quasiel_mag} and \eqref{eq:quasimag_el} with normalized,
    nondimensionalized wave profiles from SCM (solid) for two
    dispersion curves. (a) A quasi-elastic wave with $n=0$ and an
    $\mathcal{O}(\epsilon)$ magnetic component. (b) A quasi-magnetic
    wave with $j=1$ and an $\mathcal{O}(\epsilon)$ elastic
    component.}
    \label{fig:nonresonant_crossings}
\end{figure}

We compare the approximate dispersion curve predictions
\eqref{eq:el_correction} and \eqref{eq:mag_correction} with results
from the SCM code for the system \eqref{eq:54} in the top panel of
Fig.~\ref{fig:nonresonant_crossings}. Due to the higher order
correction calculated above, the actual intersection of the dispersion
curves is not equal to the intersection of the dispersion curves in
the absence of coupling. The asymptotic prediction of this new
intersection also shows excellent agreement with the SCM calculations.

We also compare the wave profiles from SCM with the asymptotic
predictions \eqref{eq:quasiel_mag} and \eqref{eq:quasimag_el} in the
bottom panel of Fig.~\ref{fig:nonresonant_crossings}. 
The asymptotic predictions show excellent agreement with numerical
calculations.

\subsection{Resonant interactions}
\label{sec:reson-inter}

Next, we examine the interactions near intersections of elastic and
magnetic dispersion curves, i.e. when the conditions
\eqref{eq:el_resonance} and \eqref{eq:mag_resonance} are met. We
denote uncoupled dispersion intersection points as
$(\tilde{k}_*,\tilde{\omega}_*)$, which are given by
$\tilde{\omega}_*=\tilde{\omega}_j(\tilde{k}_*)=\tilde{\omega}_n(\tilde{k}_*)$
for some $j,n$ in \eqref{eq:uncoupled_dispersion}, with $\tilde{k}_*$
the corresponding wave number. From \eqref{eq:el_uncoup_disp} and
\eqref{eq:mag_uncoup_disp}, we have \begin{subequations}
  \label{eq:int_point}
  \begin{align}
    \tilde{k}_*^2 &= -\tilde{H}+
                    \frac{1}{2} -\tilde{\xi}_j^2 +\frac{1}{2}
                    \sqrt{1+4(\tilde{G}^2\tilde{\xi}_n^2-
                    \tilde{\xi}_j^2 -\tilde{H})}\,, \\
    \tilde{\omega}_*^2&=\tilde{k}_*^2 + \tilde{G}^2 \tilde{\xi}_n^2 =
                        (\tilde{H}+\tilde{k}_*^2+\tilde{\xi}_j^2)^2,
  \end{align}
\end{subequations}
where we have made a sign choice for $\tilde{k}_*^2$ in order to have
real solutions for $\tilde{k}_*$.  We expand around the intersection
point $(\tilde{k}_*,\tilde{\omega}_*)$ for $0 < \epsilon \ll 1$ as
\begin{equation}
  \label{eq:as_expand}
  \begin{split}
    \tilde{k} &\sim \tilde{k}_* + \epsilon \Delta , \qquad \, \
                \tilde{A}_n(\tilde{x}_3) \sim
                \tilde{A}_n^{(0)}(\tilde{x}_3) + \epsilon
                \tilde{A}_n^{(1)}(\tilde{x}_3),\\ \tilde{\omega}
              &\sim \tilde{\omega}_* +  \epsilon f(\Delta), \quad
                \tilde{m}_j(\tilde{x}_3) \sim
                \tilde{m}_j^{(0)}(\tilde{x}_3) + \epsilon
                \tilde{m}_j^{(1)}(\tilde{x}_3). 
  \end{split}
\end{equation}
One key difference from the previous expansions \eqref{eq:1} and
\eqref{eq:9} is that an expansion for the wavenumber $\tilde{k}$ is
also included, i.e., we are only considering the dispersion in a
neighborhood of the intersection point
$(\tilde{k}_*,\tilde{\omega}_*)$. The variables $\Delta$ and
$f(\Delta)$ represent small changes in the wavenumber and frequency,
respectively. The zeroth order solutions $\tilde{m}_j^{(0)}$ and
$\tilde{A}_n^{(0)}$ again correspond to the solutions found above in
\eqref{eq:fund_soln} but they are now both included in the leading
order asymptotic expansion \eqref{eq:as_expand}.

At $\mathcal{O}(\epsilon)$, upon simplification using the
relationships between $\tilde{\omega}_*$ and $\tilde{k}_*$ given in
\eqref{eq:int_point} as well as \eqref{eq:fund_soln}, the system
\eqref{eq:64} becomes
\begin{widetext}
  \begin{subequations}
    \label{eq:80v2}
    \begin{align}
      \label{eq:80v2a}
      (\tilde{\omega}_*^2-\tilde{k}_*^2)\tilde{A}_n^{(1)}
      +\tilde{\omega}_* \tilde{G}^2 (\tilde{A}_n^{(1)})''
      &= \tilde{k}_* \tilde{m}_j^{(0)} +
        \tilde{A}_n^{(0)}\left[2\tilde{k}_*  \Delta -2
        \tilde{\omega}_* f\right], \\ 
      \label{eq:80v2b}
      (\tilde{\omega}_*^2-(\tilde{H}+\tilde{k}_*^2)^2)\tilde{m}_j^{(1)}+2
      (\tilde{H}+\tilde{k}_*^2) (\tilde{m}_j^{(1)})'' -
      (\tilde{m}_j^{(1)})'''' &= \left(4
                                \tilde{k}_*\tilde{\omega}_*\Delta- 2
                                \tilde{\omega}_* f \right)
                                \tilde{m}_j^{(0)} +
                                \tilde{\omega}_*\tilde{k}_* \tilde{A}_n^{(0)}.
    \end{align}
  \end{subequations}
  In order for the above system \eqref{eq:80v2} to be solvable, we
  require that the right hand side be orthogonal to
  $[\cos(\tilde{\xi}_n \tilde{x}_3), \, \cos(\tilde{\xi}_j
  \tilde{x}_3) ]^T$, i.e., the following inner products hold
  \begin{subequations}
    \label{eq:inner_prod}
    \begin{align}
      \label{eq:inner_prod_a}
      \int_0^{\tilde{d}} [  \tilde{k}_* \tilde{a}_{{\rm m},j}
      \cos(\tilde{\xi}_j \tilde{x}_3) + (2\tilde{k}_*  \Delta -2
      \tilde{\omega}_* f)\tilde{a}_{{\rm el},n}\cos(\tilde{\xi}_n
      \tilde{x}_3)  ] \cos(\tilde{\xi}_n \tilde{x}_3) \,
      d\tilde{x}_3 &=0, \\ 
      \int_0^{\tilde{d}} [(4 \tilde{k}_*\tilde{\omega}_*\Delta-
      2 \tilde{\omega}_* f ) \tilde{a}_{{\rm m},j} \cos(\tilde{\xi}_j
      \tilde{x}_3) +  \tilde{\omega}_*\tilde{k}_* \tilde{a}_{{\rm el},n}
      \cos(\tilde{\xi}_n \tilde{x}_3) ] \cos(\tilde{\xi}_j
      \tilde{x}_3) \, d\tilde{x}_3 &= 0.
    \end{align}
  \end{subequations}
\end{widetext}

The intersections of resonant modes occur where $n=j$ in
\eqref{eq:inner_prod}. Then the above integrals imply
\begin{equation}
  \label{eq:81}
  \begin{bmatrix}\tilde{k}_*  & 2(\tilde{k}_*  \Delta - 
                                \tilde{\omega}_* f) \\ 
    4 \tilde{k}_*\Delta- 2 f   & \tilde{k}_*  \end{bmatrix} 
  \begin{bmatrix} \tilde{a}_{{\rm m},n} \\\ \tilde{a}_{{\rm el},n} \end{bmatrix}=0. 
\end{equation}
In order to have a nonzero solution for the coefficients
$\tilde{a}_{{\rm m},n}$ and $\tilde{a}_{{\rm el},n}$, the matrix in \eqref{eq:81} must
be singular. Thus, the determinant must equal zero, yielding two
relationships between the change in wavenumber $\Delta$ and frequency
$f$,
\begin{equation}
  \label{eq:as_comp}
  f_\pm(\Delta) = \frac{\tilde{k}_*}{2\tilde{\omega}_*} \left(
    \Delta+2\Delta\tilde{\omega}_*  \pm
    \sqrt{\Delta^2(2\tilde{\omega}_*-1)^2+\tilde{\omega}_*} \right). 
\end{equation}
The two branches of $f_\pm$ in \eqref{eq:as_comp} correspond to the
two branches of the anticrossing.

The corresponding null space of the singular matrix yields the
relative sizes of $\tilde{a}_{{\rm m},n}$ and $\tilde{a}_{{\rm el},n}$
\begin{equation}
  \label{eq:rel_sizes}
  \left(\Delta  - 2\Delta \tilde{\omega}_* \mp
    \sqrt{\Delta^2(2\tilde{\omega}_*-1)^2+\tilde{\omega}_*}\right)\tilde{a}_{{\rm
      el},n} = \tilde{a}_{{\rm m},n}.
\end{equation}

Analyzing \eqref{eq:rel_sizes} reveals that for $f_+$, as
$\Delta \to -\infty$, the magnitude of the elastic displacement
$\tilde{a}_{{\rm el},n}$ grows large in comparison to magnetism
$\tilde{a}_{{\rm m},n}$. As $\Delta \to +\infty$,
$\tilde{a}_{{\rm m},n}$ grows large in comparison to
$\tilde{a}_{{\rm el},n}$. These behaviors are reversed for $f_-$. This
implies that the top anticrossing curve transitions from quasi-elastic
to quasi-magnetic, while the bottom curve transitions from
quasi-magnetic to quasi-elastic. A plot comparing the predicted
quasi-elastic and quasi-magnetic dispersion curves with the SCM
applied to eqs.~\eqref{eq:54} is provided in Fig.~\ref{fig:as_comp}.

Since $f(0)=\pm \frac{\tilde{k}_*}{2\sqrt{\tilde{\omega}_*}}$, the
anticrossing frequency gap has width
$\epsilon \tilde{k}_* /\sqrt{\tilde{\omega}_*}$.  The dimensional
frequency gap is eq.~\eqref{eq:3}.  This implies that for
anticrossings in a layered material, the gap width depends not only on
the coupling strength but also on which intersection is considered in
\eqref{eq:int_point}. For anticrossings at lower wavenumbers and
higher frequencies, the gap width is predicted to be smaller.

For nonresonant interactions, i.e. when $n \neq j$, then the integral
of $\cos(\tilde{\xi}_n \tilde{x}_3)\cos(\tilde{\xi}_j \tilde{x}_3)$ in
\eqref{eq:inner_prod} is zero. What remains is the following system of
equations
\begin{equation}
  \label{eq:83}
  \tilde{k}_* \Delta - \tilde{\omega}_* f = 0, \qquad 4 \tilde{k}_*
  \Delta - 2f = 0.
\end{equation}
The system of equations \eqref{eq:83} is solved either by $\Delta=f=0$
or $\tilde{\omega}_*=1/2$. The second of these solutions is not
generic. The first solution implies that higher order effects are
necessary, as there is no interaction at $\mathcal{O}(\epsilon)$. This
is consistent with the earlier finding that, away from anticrossings,
i.e., for nonresonant interactions, magnetoelastic coupling only has a
$\mathcal{O}(\epsilon^2)$ effect on the dispersion curves.

To summarize, in this Appendix, we showed that finite thickness effects
lead to an infinite number of dispersion curves in a material. The
many intersections between magnetic and elastic curves can be
classified as resonant or nonresonant, depending on whether the
quasi-elastic and quasi-magnetic waves share the same order. For
exchange-dominated waves, only resonant interactions yield
anticrossings. Nonresonant interactions only display simple crossings,
which nevertheless experience an $\mathcal{O}(\epsilon^2)$ shift from
their uncoupled intersections.

\section{Validation of SCM}
\label{sec:validation-scm}

\begin{figure}
  \centering
  \includegraphics[scale=.25]{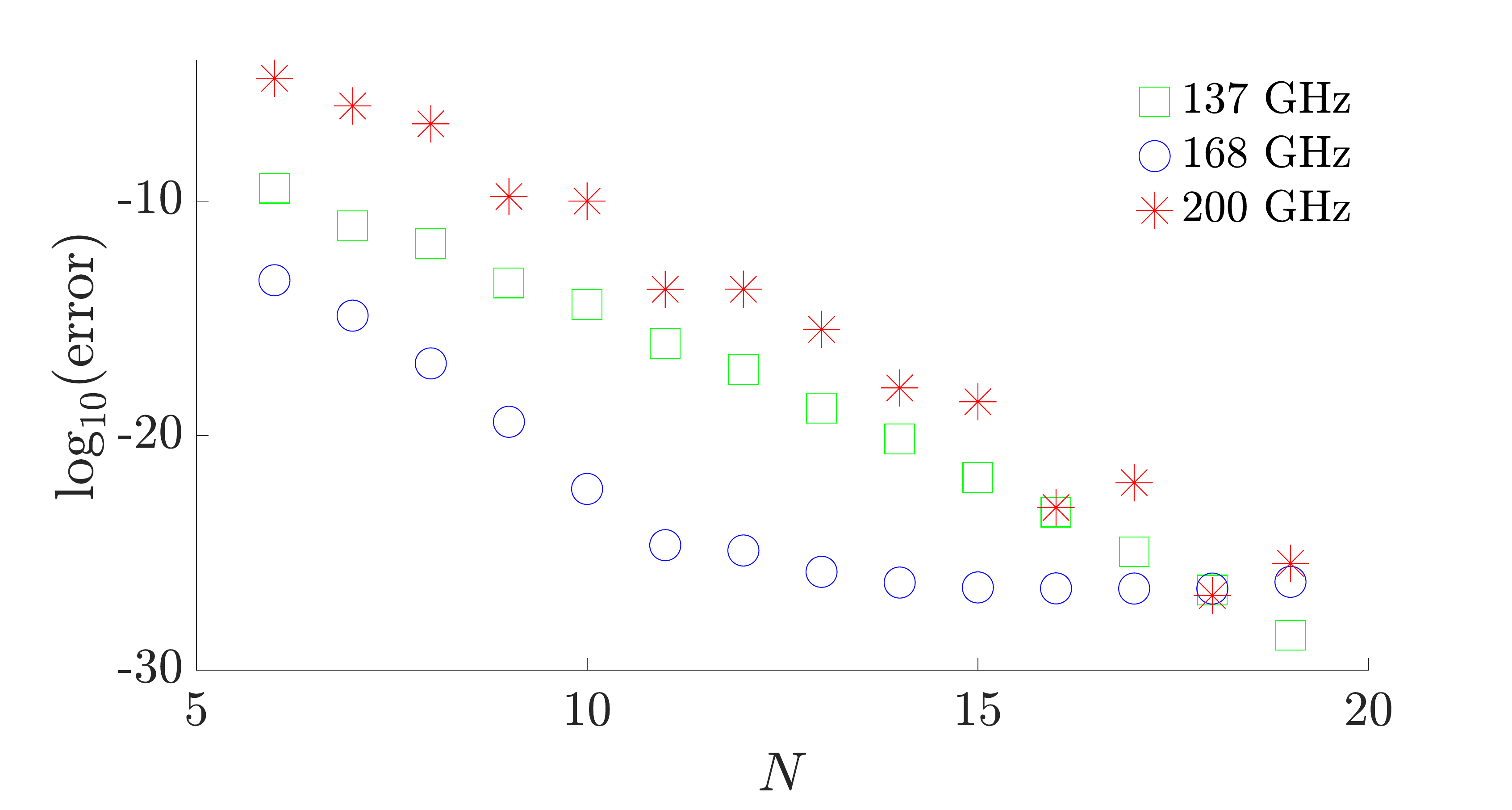}
  \caption{Convergence of frequency values for $k=0.3$ rad/nm on a
    logarithmic scale. The material studied is 30 nm of YIG on 50 nm
    of GGG. Each data point shows the relative error with the
    frequency value calculated for $N=20$. Due to calculations
    performed in quadruple precision, convergence is to $10^{-30}$,
    after which rounding errors dominate.}
    \label{fig:freq_conv}
\end{figure}

In order to validate the SCM approach, we examine the convergence of
eigenvalues (i.e. frequencies) as the discretization $N$ is
increased. In Fig.~\ref{fig:freq_conv}, we show the convergence of
three frequency values for $k=0.3$ rad/nm corresponding to three types
of waves for the YIG-GGG double layer. The three frequency values were
calculated for each value of $N$ for $6<N<20$, and then the relative
error with the frequency calculated at $N=20$ is displayed on a log
scale. Even for higher frequencies, the convergence is rapid and the
errors are small.
\begin{figure}
  \centering
  \includegraphics[scale=.25]{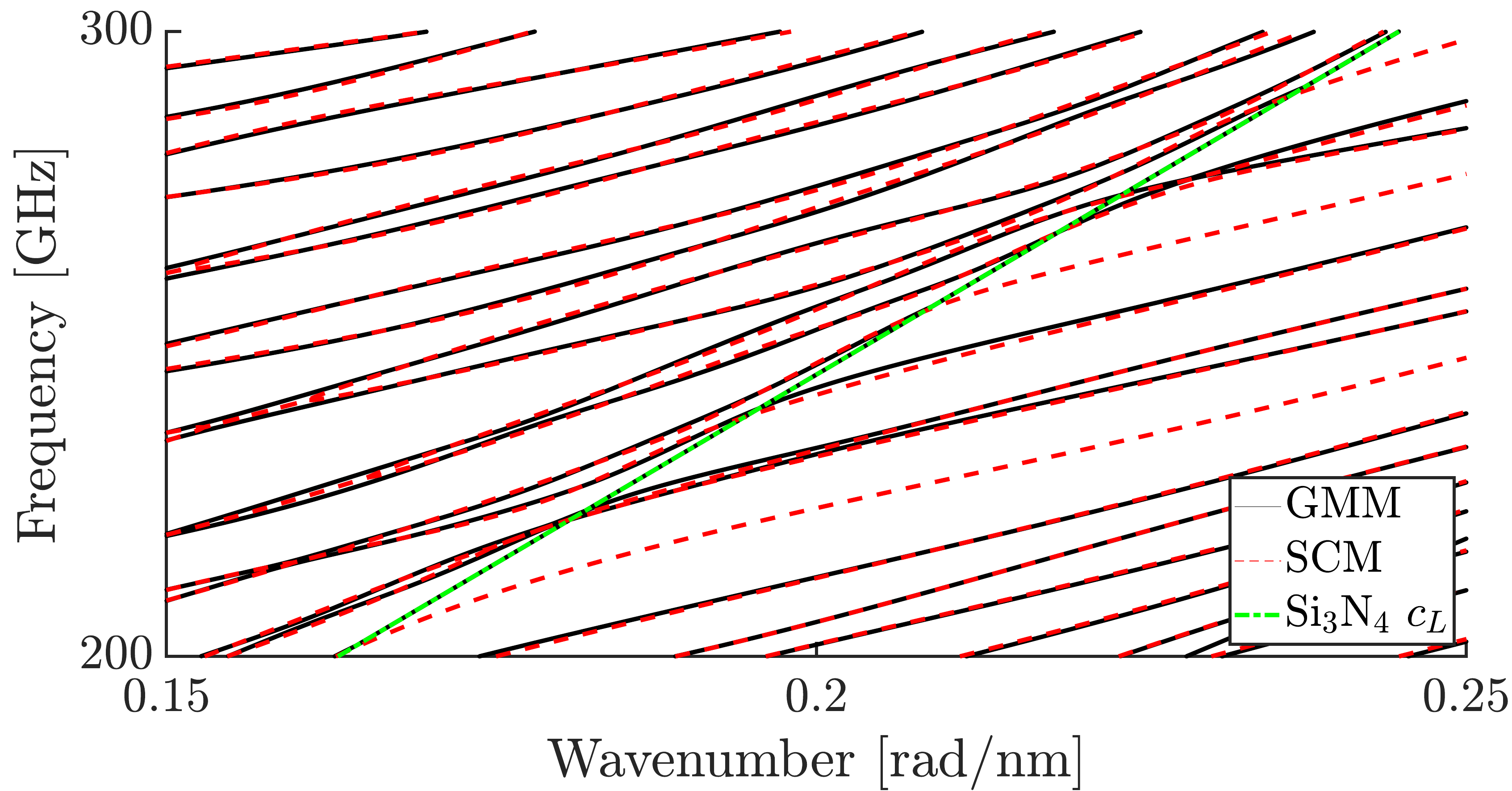}
  \caption{Comparison between SCM (dashed) and GMM (solid) methods in
    the absence of magnetism for 50 nm of Ni layered on 100 nm of
    Si$_3$N$_4$ with $N=24$. GMM incorrectly identifies the
    Si$_3$N$_4$ longitudinal speed of sound $c_L$ as a dispersion
    curve, and other GMM dispersion curves remain trapped in this
    incorrect minimum.}
    \label{fig:scm_gmm_compare}
\end{figure}

As another validation step, we compare SCM results to GMM results from
a published solver, namely the ElasticMatrix MATLAB toolbox
\cite{ramasawmy_elastic_mat}. ElasticMatrix computes dispersion curves
in a layered, purely elastic material utilizing a GMM
approach. Ignoring the magnetic coupling, we calculate elastic
dispersion curves for the Ni-Si$_3$N$_4$ double layer using our SCM
code by zeroing the coupling coefficient $B_2 = 0$ and compare with
the toolbox results.

The results of this comparison for $N=24$ are presented in
Fig.~\ref{fig:scm_gmm_compare}. For most all of the dispersion curves,
excellent agreement between the SCM and GMM approaches are
obtained. It is important to explain the discrepancies. First, the GMM
approach incorrectly identifies the silicon nitride longitudinal speed
of sound as a dispersion curve (dashed green). This has been
previously identified as a shortcoming of GMM calculations
\cite{quintanilla_2015_spectral}.
Second, some curves are only found by the SCM approach. We hypothesize
that this occurs when the GMM curve-tracing algorithm ends up trapped
by the (incorrect) speed of sound dispersion curves. Missed curves
have also been previously reported as a shortcoming of the GMM method
\cite{quintanilla_2015_spectral}.

We stress that, running on the same machine, the SCM code to generate
Fig.~\ref{fig:scm_gmm_compare} had a run time of less than ten
minutes. In contrast, the ElasticMatrix method required over ten hours
to compute nearly identical dispersion curves.

A third validation for SCM is its close agreement with the asymptotic
analysis shown in Sec.~\ref{sec:mag_asymp}.  Since the asymptotic
calculation included an analytical calculation of the uncoupled
magnetic and elastic dispersion curves in a simplified layered
material, the agreement in the vicinity of the anticrossing in
Fig.~\ref{fig:as_comp} validates that the code accurately recovers
these curves. Combined with the excellent agreement for nonresonant
crossings in Fig.~\ref{fig:nonresonant_crossings}, we have ample
evidence that the SCM reliably incorporates coupling effects.



\providecommand{\noopsort}[1]{}\providecommand{\singleletter}[1]{#1}%

\end{document}